\documentclass[10pt, sigconf, twocolumn]{acmart}
\usepackage[export]{adjustbox}
\usepackage{pbox}
\usepackage{caption,graphicx,newfloat}
\usepackage{afterpage}
\usepackage{algorithm,algpseudocode}
\usepackage{pifont}
\usepackage{amsmath}
\usepackage{soul}
\usepackage{listings}
\usepackage{empheq}
\usepackage[framemethod=tikz,xcolor=true]{mdframed}
\usepackage{etoolbox}

\usepackage{amsthm}
\usepackage{amsfonts}
\usepackage{colortbl}
\usepackage{paralist}
\usepackage{enumitem}
\usepackage{epsfig}
\usepackage{forest}
\usepackage{graphicx}
\usepackage{multicol}
\usepackage{multirow}
\usepackage{pgf}
\usepackage{pdflscape}
\usetikzlibrary{positioning}
\usepackage{rotating}
\usepackage[normalem]{ulem}
\usepackage{relsize}
\usepackage{textcomp}

\usepackage{tikz-cd}

\usepackage{wrapfig}
 \usepackage{xspace}
\usepackage{balance}
\usepackage{thmtools,thm-restate}
\usepackage[normalem]{ulem}
\usepackage{url}
\usepackage{amsmath}
\usepackage{amssymb}
\usepackage{subcaption}
\usepackage{amsthm}
\usepackage[normalem]{ulem}
\usepackage{forest}

\usepackage{xcolor}
\usepackage{listings}

\usepackage{xparse}

\renewcommand{\paragraph}[1]{\vspace{2pt} \noindent {\bf #1}}

\newcommand{\rev}[1]{{{#1}}}

\newcommand{\revtr}[1]{{{#1}}}

\newcommand{\eat}[1]{}
\newcommand{\later}[1]{}

\newcommand{\agp}[1]{\textcolor{blue}{#1}}

\newcommand{\agppapertext}[1]{{ }}
\newcommand{\agptechreport}[1]{#1}

\newcommand{\tar}[1]{\textcolor{violet}{Tarique: #1}}
\newcommand{\cut}[1]{}

\newcommand{\revisit}[1]{}

\newcommand{\papertext}[1]{}
\newcommand{\techreport}[1]{}

\newcommand{\smallcaption}[1]{{\small{#1}}}

\newcommand{\new}[1]{#1}

\newcommand{\ssr}{{\sf ShapeSearch}\xspace}
\newcommand{\ssrc}{{\sf ShapeSearch$^*$}\xspace}

\newcommand{\sq}{{\sf ShapeQuery}\xspace}
\newcommand{\sqs}{{\sf ShapeQueries}\xspace}
\newcommand{\ssg}{{\sf ShapeSegment}\xspace}
\newcommand{\ssgs}{{\sf ShapeSegments}\xspace}

\newcommand{\sgt}{{\sf SegmentTree}\xspace}

\newcommand{\pattern}{{\sf PATTERN}\xspace}
\newcommand{\paterns}{{\sf PATTERNs}\xspace}

\newcommand{\location}{{\sf LOCATION}\xspace}
\newcommand{\modifier}{{\sf MODIFIER}\xspace}
\newcommand{\iterator}{{\sf ITERATOR}\xspace}
\newcommand{\sketch}{{\sf SKETCH}\xspace}
\newcommand{\POSITION}{{\sf POSITION}\xspace}
\newcommand{\match}{{\sf MATCH}\xspace}
\newcommand{\concat}{{\sf CONCAT}\xspace}
\newcommand{\AND}{{\sf AND}\xspace}
\newcommand{\OR}{{\sf OR}\xspace}
\newcommand{\opposite}{{\sf OPPOSITE}\xspace}

\newtheorem{definition}{Definition}[section]

\newtheorem{problem}{Problem}

\newtheorem{theorem}{Theorem}[section]
\newtheorem{assumption}{Assumption}[section]

\newcounter{enum}

\newif\iftechreportcode
\techreportcodetrue

\newif\ifpapertext

\newcommand{\cmark}{\ding{51}}%
\newcommand{\xmark}{\ding{55}}%

\NewDocumentCommand{\loc}{v}{%
	\texttt{\textcolor{brown}{#1}}%
}

\NewDocumentCommand{\pat}{v}{%
	\texttt{\textcolor{olive}{#1}}%
}

\newcommand{\thetx}{$\color{olive}\theta=x$}%
\newcommand{\p}{\color{olive}p}%
\newcommand{\pd}{\color{olive}\$}%

\newcommand{\colorconcat}{$\color{teal}\otimes $}%
\newcommand{\colorand}{$\color{brown}\odot $}%
\newcommand{\coloror}{$\color{orange}\oplus$}%

\newcommand{\xs}{\color{brown}x.s}%
\newcommand{\xe}{\color{brown}x.e }%
\newcommand{\ys}{\color{brown}y.s}%
\newcommand{\ye}{\color{brown}y.e}%
\newcommand{\ld}{\color{brown}.}%
\newcommand{\vs}{\color{brown}v}%

\newcommand{\pup}{{\color{olive} p=up}}%
\newcommand{\pdown}{{\color{olive} p=down}}%
\newcommand{\pflat}{{\color{olive} p=flat}}%

\NewDocumentCommand{\md}{v}{%
	\texttt{\textcolor{purple}{#1}}%
}
\newcommand{\m}{\color{purple}m}%
\newcommand{\mgt}{\color{purple}>}%
\newcommand{\mgteq}{\color{purple}>2}%
\newcommand{\meq}{\color{purple}=}%

\newcommand{\squishlist}{
   \begin{list}{$\bullet$}
    { \setlength{\itemsep}{0pt}
      \setlength{\parsep}{2pt}
      \setlength{\topsep}{0pt}
      \setlength{\partopsep}{0pt}
      \leftmargin=25pt
\rightmargin=0pt
\labelsep=5pt
\labelwidth=10pt
\itemindent=0pt
\listparindent=0pt
\itemsep=\parsep
    }
}

\DeclareFloatingEnvironment[
  fileext=lob,
  listname={List of Boxes},
  name=Table
]{BOX}

\newenvironment{denselist}{
    \begin{list}{\small{$\bullet$}}%
    {\setlength{\itemsep}{0ex} \setlength{\topsep}{0ex}
    \setlength{\parsep}{0pt} \setlength{\itemindent}{0pt}
    \setlength{\leftmargin}{0.85em}
    \setlength{\partopsep}{0pt}}}%
    {\end{list}}

\newcommand{\squishend}{\end{list}}

\newcommand{\stitle}[1]{\vspace{0.25em}\noindent\textbf{#1}}
\newcommand{\rom}[1]
{\MakeUppercase{\romannumeral #1}}

\renewcommand\footnotetextcopyrightpermission[1]{} 

\begin{document}

\title{{\ssr}: A Flexible and Efficient System for Shape-based Exploration of Trendlines}
\author{{\large Tarique Siddiqui\textsuperscript{1}, Zesheng Wang\textsuperscript{1}, Paul Luh\textsuperscript{1},  Karrie Karahalios\textsuperscript{1}, Aditya G. Parameswaran\textsuperscript{2}} \\
	{\large \textsuperscript{1}University of Illinois (UIUC), \textsuperscript{2}UC Berkeley} \\
	{\large \{tsiddiq2, zwang180, luh2, kkarahal\}@illinois.edu}, \large adityagp@berkeley.edu}


\begin{abstract}
Identifying trendline visualizations with desired patterns 
is a common task during data exploration. 
Existing visual analytics tools offer 
limited \emph{flexibility}, \emph{expressiveness}, and \emph{scalability}
for such tasks, especially when the pattern of interest
is under-specified and approximate. 
We propose \ssr, an efficient and flexible pattern-searching tool, 
that enables the search for desired patterns 
via multiple mechanisms: sketch, natural-language, and visual regular expressions. 
We develop a novel \emph{shape querying algebra}, 
with a minimal set of primitives and operators 
that can express a wide variety of \ssr queries, 
and design a natural-language and regex-based parser 
to translate user queries 
to the algebraic representation. 
To execute these queries within interactive response times, 
\ssr uses a fast shape algebra execution engine 
with query-aware optimizations, 
and perceptually-aware scoring methodologies. 
We present a thorough evaluation of the system, 
including a user study, 
a case study involving genomics data analysis, 
as well as performance experiments, comparing against state-of-the-art trendline 
shape matching approaches---that together demonstrate the usability and scalability of \ssr.
\end{abstract}

\maketitle
\setcounter{page}{1}
\setcounter{section}{0}
\renewcommand\thefigure{\arabic{figure}}
\setcounter{figure}{0}


\vspace{-5pt}
\section{Introduction}
\label{sec:intro}

Identifying patterns in trendlines or line charts 
is an integral part of data exploration---routinely performed 
by domain experts to make sense of their datasets, 
gain new insights, and validate their hypotheses. 
For example, clinical data analysts 
examine trends of health indicators 
such as temperature and heart-rate for diagnosis 
of medical conditions~\cite{gotz2014methodology}; 
astronomers study the variation in properties of 
galaxies over time to understand the history and 
makeup of the Universe~\cite{morganson2018dark}; 
biologists analyze gene expression patterns 
over time to study biological processes~\cite{wagner2005genome,lee2019you}; 
and financial analysts study trends
in stock prices to predict future behavior~\cite{ge1998pattern}. 
Due to the lack of extensive programming experience, 
these domain experts 
typically perform manual exploration, tediously examining trendlines
one at a time until they find ones that match their desired shape or pattern, 
e.g., gene expressions that rise and then become stable.

Recent work has proposed tools that let users 
interactively search for 
desired patterns~\cite{zenvisagevldb, qetchchi, timesearcher,googlecorrelate}. 
However, as we will discuss below, 
these tools expect users to search in highly {\em constrained} ways,
and, in addition, are {\em overly rigid} in how they assess a match.
Most tools expect users to specify 
a complete and exact trendline as input usually by sketching it on a canvas, 
followed by
computing distances between this exact trendline 
and several candidate trendlines to identify matches.
As a result, these tools 
are unable to support search when the desired shape is 
{\em under-specified or approximate},
e.g., finding stocks whose prices are decreasing for some time, 
followed by a sharp rise, with the position and intensity of 
movements being left unspecified, or when the desired shape is {\em complex},
e.g., finding gene expression profiles where there is an unspecified
number of peaks and valleys followed by a flattening out. 
Some data mining tools provide the ability to search for
patterns in time series, e.g.,~\cite{psaila1995querying,garofalakis1999spirit},
but require heavy precomputation,
limiting ad-hoc exploration, in addition to suffering from the same
limitations in flexibility as the visualization tools. 
Yet another alternative for domain experts with programming expertise
is to write code to perform this flexible match, but writing code for
each new use-case, followed by manual optimization is often as
tedious as manual searching of visualizations to find patterns.

\begin{figure}
	 \vspace{-25pt}
	\centering\includegraphics[width=\columnwidth]{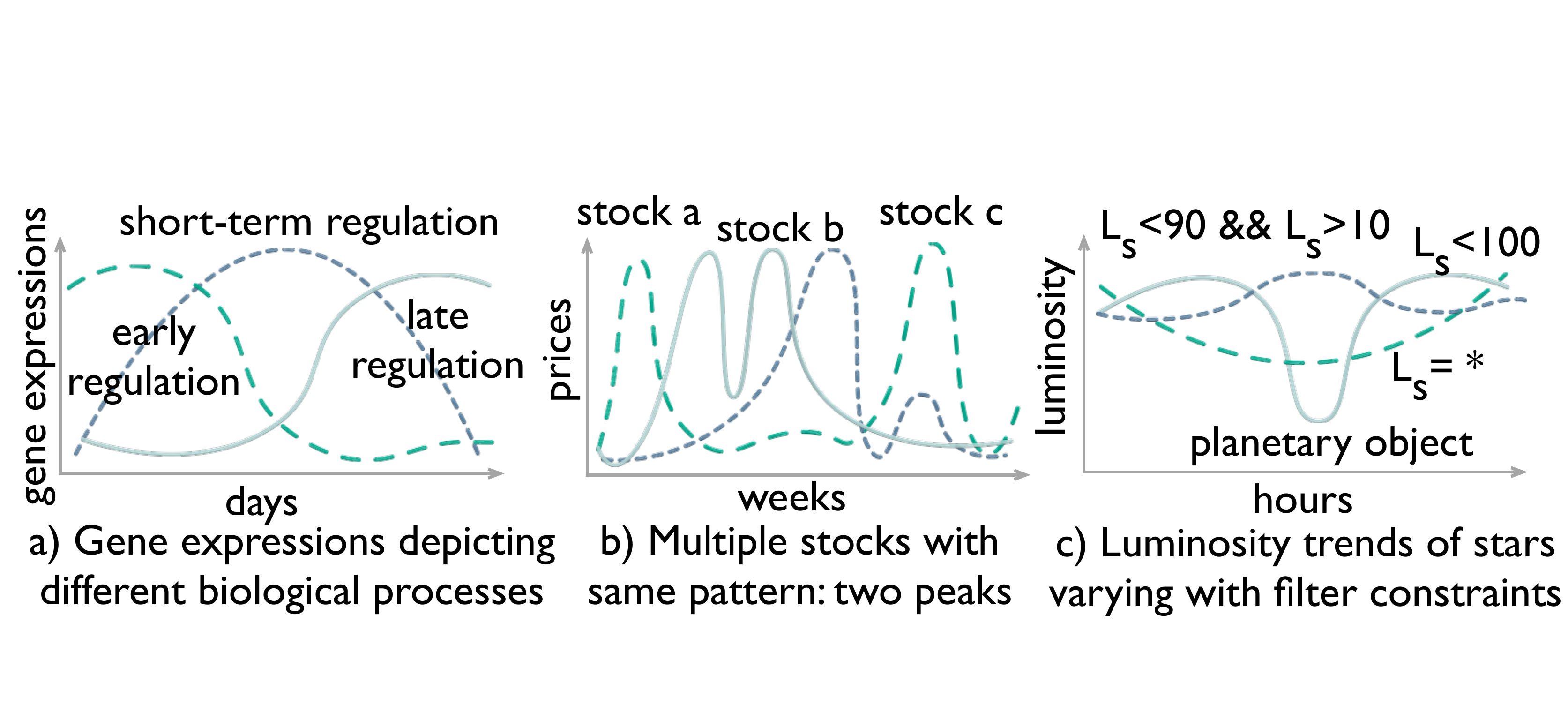}
	\vspace{-35.5pt}
	\caption{\smallcaption{Shapes characterizing real world phenomena}}
	\label{fig:motiv_example}
	\vspace{-15.5pt}
\end{figure}

\eat{
\begin{figure*}[!t]
\papertext{\vspace{-40pt}}
\centering\includegraphics[width=.85\textwidth]{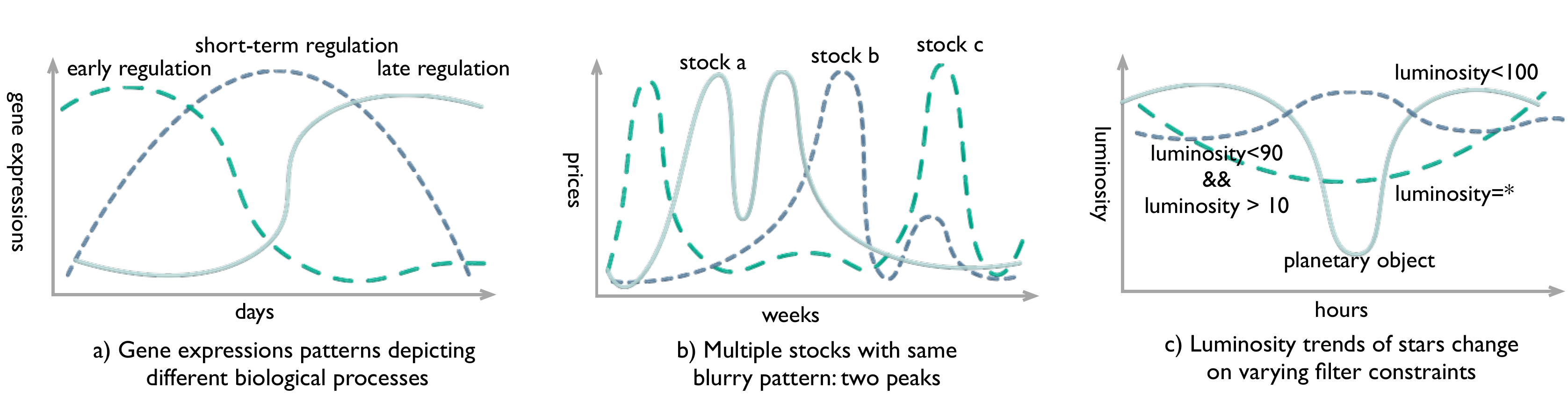}
\vspace{-10.5pt}
\caption{\smallcaption{Shapes characterizing various real world phenomena}}
\label{fig:motiv_example}
\vspace{-10.5pt}
\end{figure*}
}

We present \ssr, a visual data exploration system 
that supports multiple novel mechanisms to 
express and effortlessly search for 
desired patterns in trendlines. Before 
describing \ssr, we first characterize 
typical trendline pattern-based queries.

\vspace{-5pt}
\subsection{Characterizing Shape Queries} 
\label{sec:formativestudy}
\rev{The design of \ssr has been motivated by case studies and use-cases 
from domains such as genomics, astronomy, battery science, and finance, using a process similar to our earlier work~\cite{lee2019you}.}
We also collected a corpus of about $250$
natural language queries via Mechanical Turk (mturk), 
where we asked crowd workers to describe patterns in 
trendline visualizations collected from real world datasets\footnote{Described in more detail in Appendix~\ref{sec:mturk}.}. 
We highlight the key characteristics of
pattern matching tasks, 
based on our discussions with domain experts and 
analysis of mturk queries below.

\stitle{Fuzzy Matching.} 
Domain experts typically search for 
patterns (i) that are \emph{approximate}, 
and are often not interested
in the specific details or local fluctuations
as much
as the overall shape, and 
(ii) they often \emph{do not}
specify or even know the exact 
location of the occurrence of patterns.
For example, biologists
routinely look for structural changes in 
gene expression, e.g., rising and falling 
at different times (Figure~\ref{fig:motiv_example}a). 
Structural changes characterize internal 
biological processes such as the 
cell cycle or circadian rhythms, 
or external perturbation, 
such as the influence of a drug 
or presence of a disease. 
Similarly, many crowd workers tend to describe
 trendlines using high level patterns such as 
 \emph{increasing and then decreasing}, 
 without being precise about locations and/or 
 features of the changes.


\stitle{Combination of Multiple Simple Patterns.} 
We notice that both domain experts as well as crowd workers 
often describe complex patterns using a 
\emph{combination of multiple simple ones}. 
Each individual pattern is typically described 
using words such as "increasing", "stable", or "falling", 
which are easy to state in natural language
but hard to specify using existing query languages. 
Moreover, pattern matching tasks in many domains 
often go beyond 
finding a sequence of patterns, 
requiring arbitrary combinations, e.g., 
disjunction, conjunction, or quantification, 
with varying location or width constraints.  
Examples include finding stocks with at least 
$2$ peaks within a span of $6$ months, e.g., the so-called "double/triple top" patterns 
that indicate future downtrends~\cite{investopedia}, 
or finding cities where the temperature 
rises from November to January and falls during 
May to July, such as Sydney. 


\stitle{Ad hoc and Interactive.} 
Pattern-based queries 
are often defined {\em on-the-fly} during analysis, 
based on other patterns observed. 
For instance, biologists often search 
for a pattern in a group of genes 
similar to a pattern recently discovered 
in another group~\cite{lee2019you}. 
Similarly, astronomers monitor the shape of the luminosity trends 
of stars over time  
to search for and 
characterize new planetary objects (Figure~\ref{fig:motiv_example}c).
For example, a dip in brightness 
often indicates a planetary object 
passing between the star and the telescope.
\revtr{
In order to limit comparison of patterns over similar duration (i.e., the X axis) or over value ranges (i.e, the Y axis), it is common to apply constraints while pattern matching. Examples include searching for changes in buying and selling patterns of stock or house prices in a specific range or duration. As such, some tools, e.g., TimeSearcher~\cite{timesearcher}, allow interactive specification of constraints, however the pattern matching is still precise or value-based.
}

\vspace{-8pt}
\subsection{Our Approach}
To  satisfy the aforementioned characteristics, 
\ssr makes three contributions.

\stitle{(a)} \ssr incorporates an expressive 
\emph{shape query algebra} that abstracts 
key shape-based primitives and operators
for expressing a variety of desired trendline patterns. 
The most powerful feature of this algebra 
is its capability for ``fuzzy'' matching, 
allowing approximate and inexact pattern specification, 
without compromising on the needs of occasional precise queries.
We developed this algebra after 
discussions with domain experts, 
as well as studying mturk pattern queries, 
as mentioned earlier. 

\stitle{(b)} Unfortunately, na\"ively executing these 
fuzzy queries is extremely slow, 
requiring an expensive evaluation of 
all possible ways of matching each candidate trendline 
to the query 
to select the best one. 
We propose a dynamic programming-based optimal algorithm
that reuses computation  
to provide substantial speed-ups, 
and show that even this algorithm 
can be prohibitively slow for interactive ad-hoc exploration. 
We then develop a novel perceptually-aware bottom-up algorithm 
that incrementally prunes the search space 
based on patterns specified in the query, 
providing a  $\textbf{40}\times$ speedup 
with over 85\% accuracy, compared to the optimal approach.

\begin{table}[t]
\vspace{-5pt}
\caption{\smallcaption{Comparison between specification mechanisms\label{tab:spec-comparison}}}
\vspace{-12pt}
\resizebox{0.93\columnwidth}{!}{%
\begin{tabular}{l|lll}

 Mechanism & Intuitiveness  & Control  & Expressiveness  \\ \hline
 Natural language & high  & low  & high \\
 Sketch & high  & high & low  \\
 Regex & low  & high  & high 
\end{tabular}%
}
\vspace{-16pt}
\end{table}

\stitle{(c)} Finally, to accommodate a range of needs
without sacrificing the expressiveness of the algebra, 
\ssr supports three query specification mechanisms (Table~\ref{tab:spec-comparison}): 
sketching on a canvas, natural language, and regular expressions (regex for short). 
All specification mechanisms are translated 
to the same shape query algebra representation, 
and can be used interchangeably, 
as user needs evolve.

Next, we explain how a user interacts with \ssr. 


\begin{figure*}	
	\hspace{-0.2cm}
	\begin{subfigure}{0.60\textwidth}
	\vspace{-15pt}
		\centerline {
	\hbox{\resizebox{\columnwidth}{!}{\includegraphics{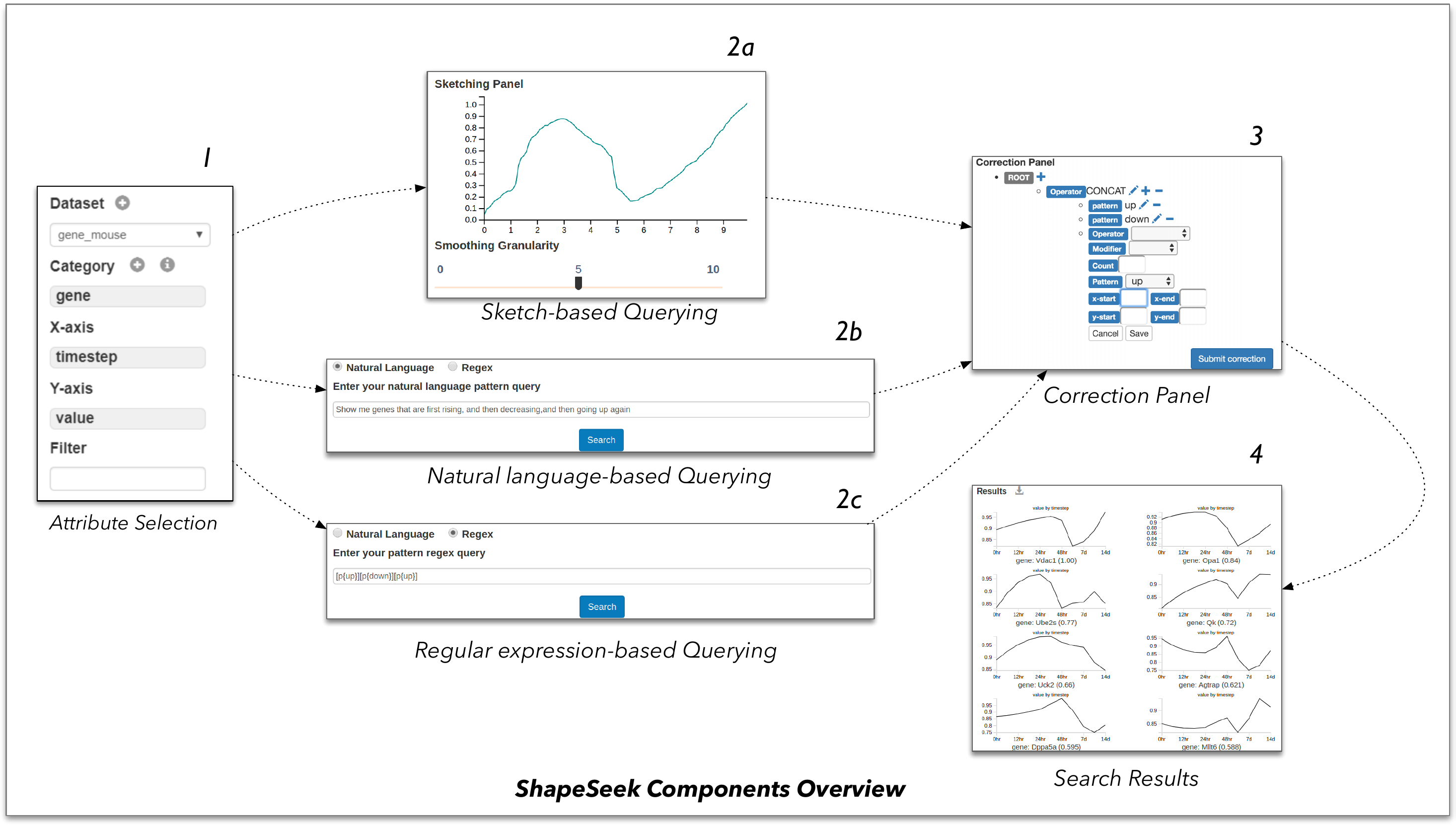}}}}
	\vspace{-2.5pt}
	\caption{\smallcaption{\ssr Interface}}
	\label{fig:interface}
	\end{subfigure}
	\hspace{0.2cm}
	\begin{subfigure}{0.34\textwidth}
	\begin{subfigure}{0.89\textwidth}
		\begin{center}
		\label{tab:algebra}
		\vspace{-15pt}
		\resizebox{\columnwidth}{!}{%
			\begin{tabular}{|l l l |} 
				\hline
				Symbol & Name & Type \\ [0.5ex] 
				\hline
				\xs & {\sf START X VALUE} & Location Sub-Primitive  \\
				\ys & {\sf START Y VALUE} & Location Sub-Primitive  \\
				\xe & {\sf END X VALUE} & Location Sub-Primitive \\
				\ye & {\sf END Y VALUE} & Location Sub-Primitive  \\ 
				\vs & \sketch & Location Sub-Primitive  \\	
				\p  & \pattern & Primitive  \\
				
				\colorconcat  & \concat & Operator   \\
				\colorand & \AND & Operator  \\
				\coloror & \OR & Operator   \\
				\hline
			\end{tabular}%
		}
		\vspace{-5pt}
	\caption {\smallcaption{Algebra Primitives and Operations}}
	     \label{tab:algebra}
	\end{center}
	\end{subfigure}
	\begin{subfigure}{\textwidth}

		\centering
		\resizebox{0.55\columnwidth}{!}{%
			\begin{tikzpicture}[node distance=2cm]
			\node (A) at (1.2, 3) {\colorconcat};
			\node (B) at (0, 2) {[\pup]};
			\node (C) at (1.2, 2) {[\pdown]};
			\node (D) at (2.5, 2) {\coloror};
			\node (E) at (1.2, 1) {\colorconcat};
			\node (F) at (3, 1) {[\pup]};
			\node (G) at (0.5, 0) {[\pup]};
			\node (H) at (1.9, 0) {[\pdown]};
			\draw[thick,->] (A) -- (B);
			\draw[thick,->] (A) -- (C);
			\draw[thick,->] (A) -- (D);
			\draw[thick,->] (D) -- (E);
			\draw[thick,->] (D) -- (F);
			\draw[thick,->] (E) -- (G);
			\draw[thick,->] (E) -- (H);
			\end{tikzpicture}%
		}
		\vspace{-3pt}
		\caption{\sq AST}
		\label{fig:ast}
	\end{subfigure}

	\end{subfigure}
	\vspace{-8pt}
	\caption{ \smallcaption{a) \ssr Interface, consisting of six components. Box 1: Data upload, attributes selection, and applying filter constraints; Box 2: Query specification---Box 2a:
			Sketching canvas, Box 2b: Natural language query interface, and 
			Box 2c: Regular expression interface; Box 3: Correction panel; and Box 4) Results panel. b) Primitives and Operators in  ShapeQuery. c) Abstract tree representation of \sq [{\color{olive} p=up}]\colorconcat 
			[{\color{olive}p=down}]\colorconcat(([{\color{olive} p=up}]\colorconcat[{\color{olive} p=down}])\coloror[{\color{olive} p=flat}])} }
	\vspace{-10pt}
	\label{fig:components}	
\end{figure*}

\eat{
\begin{figure*}
	\vspace{-25pt}
	\centerline {
		\hbox{\resizebox{0.65\textwidth}{!}{\includegraphics{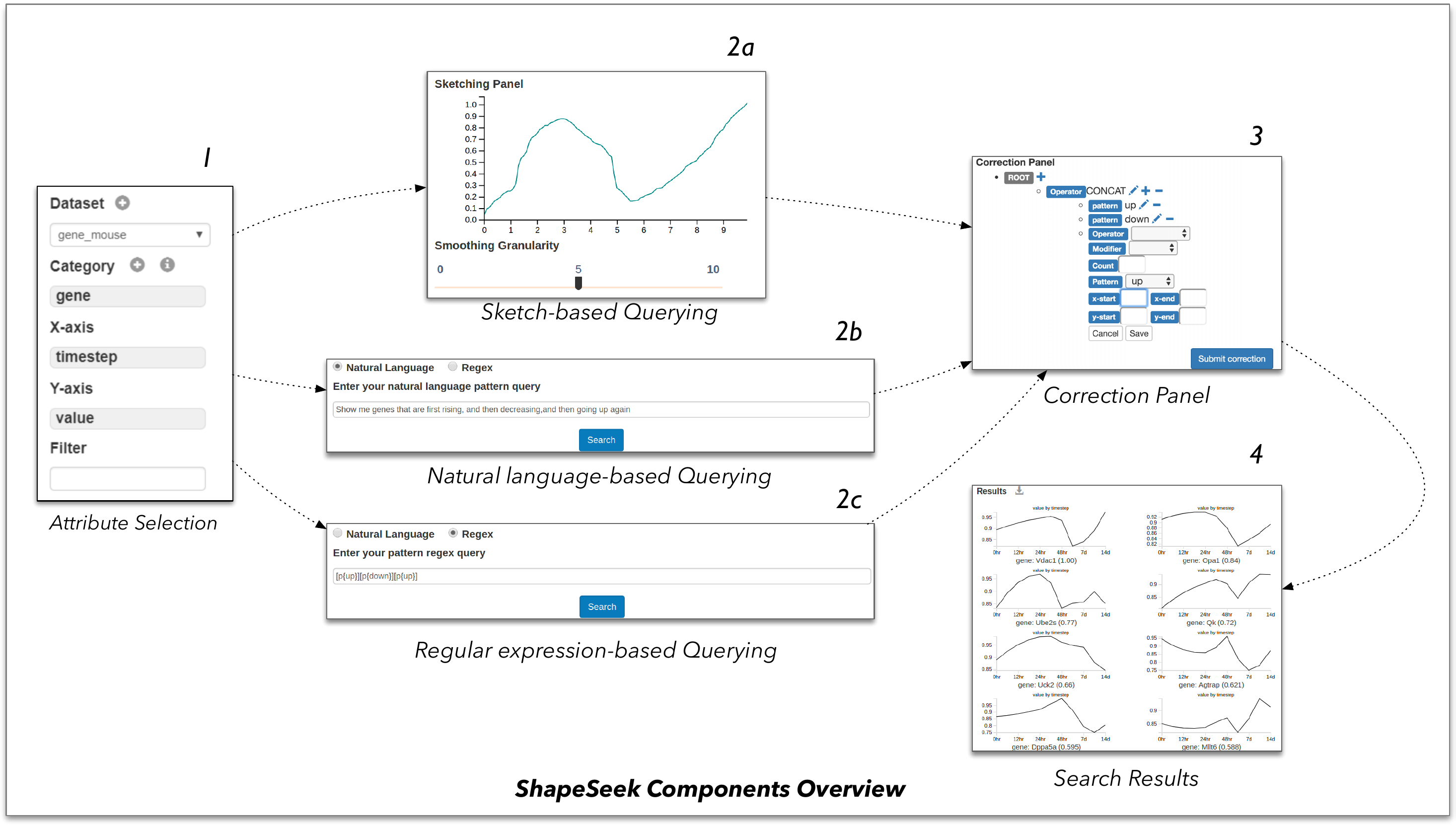}}}}
	\vspace{-2.5pt}
	\caption{\smallcaption{\ssr Interface, consisting of six components. 1) Data upload, attributes selection, and applying filter constraints 2) Query specification: 2a)
			Sketching canvas 2b) Natural language query interface, and 2c) Regular expression interface, 3) Correction panel, and 4) Results panel}}
	\vspace{-15.pt}
	\label{fig:interface}
\end{figure*}
}

\vspace{-8pt}
\subsection{\ssr System Overview} 
Figure~\ref{fig:interface} depicts the \ssr interface, 
with an example query on genomics data. 
Here, a user wants to search for genes 
that get suppressed due to the influence of a drug, 
with a specific shape in their gene expression---first 
rising, then going down, and finally rising again---with 
three patterns: up, down, and up, in a sequence. 
To search for this shape, 
the user first loads the dataset~\cite{bult2008mouse}, 
via a form (Figure~\ref{fig:interface} box 1), 
and then selects the space of trendline visualizations 
to explore by setting parameters: $x$ axis as time, 
$y$ axis as expression values, 
and category/$z$ axis  as gene. 
\ssr generates a trendline visualization for each unique value of the $z$ axis.  
Thus, the $z$ axis defines the space of visualizations 
over which we match the shape. 
Once the data is loaded, 
the user can leverage three mechanisms 
for shape query specification:

\stitle{Sketching on Canvas}.  
By drawing the desired shape as a 
sketch on the canvas (Figure~\ref{fig:interface} box 2a), 
the user can search for trendlines
that \emph{precisely} match this sketch, using a distance 
measure such as Euclidean distance or Dynamic Time Warping~\cite{dtw}.
\ssr outputs visualizations 
similar to the drawn sketch 
in the results panel (Figure~\ref{fig:interface} box 4).

\stitle{Natural Language (NL)}.
For searching for approximate pattern matches, 
users can use natural language.
For instance, in Figure~\ref{fig:interface} box 2b, the desired genomics shape 
can be expressed as \emph{``show me genes that are rising, 
then going down, and then increasing''}. 
Similarly, scientists analyzing cosmological data can 
search for supernovae (bright stellar explosions) using 
\emph{``find me objects with a sharp peak in  luminosity''}. 

\stitle{Regular Expression (regex)}. For queries that involve complex combinations of patterns 
that are difficult to express using natural language or sketch, 
the user can issue a regular expression-like query that directly maps to the internal \sq algebraic 
representation, consisting of \ssr primitives and operations. \rev{While sketch is typically used for precise matching, \ssr also allows approximate matching via sketch by constructing a regex from a sketch\agppapertext{~\cite{techreport}}\agptechreport{(see Appendix~\ref{sec:sketchapprox})}.}


The \ssr back-end parses and translates all queries into the \sq algebra before execution.  
For translating natural language queries, \ssr supports a sophisticated parser 
that uses a mix of learning and rules for resolving syntactic and semantic ambiguities. 
After translation, the backend forwards the regex representation 
of the query to the user for validation or correction (Figure~\ref{fig:interface} Box 3). 
The validated query is finally optimized and executed, 
and the top visualizations that best match the \sq are presented 
in the results panel (Figure~\ref{fig:interface} Box 4).
	
 \stitle{Paper Outline.} We explain the three key components of \ssr in the following sections. 
 In Section~\ref{sec:algebra}, we give an overview of the \sq algebra, along with its primitives and operators. 
 In Section~\ref{sec:fuzzychallenge}, we discuss the challenges in executing fuzzy shape queries and how we make \ssr scale to large collections of trendlines. 
 We briefly explain the natural language translation in Section~\ref{sec:translation}. 
 We describe our performance experiments evaluating the efficiency and accuracy of the \ssr pattern execution engine in Section~\ref{sec:exp}. 
 We present a user study in Section~\ref{sec:userstudy} and a genomics case study in Section~\ref{sec:casestudy}, 
 evaluating the expressiveness, effectiveness, and usability of \ssr. 
 \agppapertext{We provide additional details on the front-end, back-end, and usage scenarios of \ssr 
 in our technical report~\cite{techreport}.} 
 We presented an early version of \ssr in a demo paper~\cite{shapesearchdemo}.

\vspace{-5pt}
\section{ShapeQuery Algebra}
\label{sec:algebra}

We give an overview of \sq, a structured query algebra, 
motivated from use-cases in real domains as well as our 
analysis of the crowdsourced pattern queries. 


\stitle{Overview.} The \sq algebra 
consists of a minimal set of primitives and operators 
for declaratively expressing a rich variety of patterns, while 
supporting the three characteristics of pattern-matching tasks described in the introduction. 
At a high level, a \sq represents a {\em shape} as a combination of multiple {\em simple patterns}. 
A simple pattern can either be precise with specific location constraints, 
e.g., matching $y=x$ between $x=2$ to $x=6$, 
or fuzzy, e.g., roughly increasing, where the notion of the pattern 
is approximate and its location unspecified. 
Each simple pattern along with its precise or imprecise 
constraints is called a {\em \ssg}. 
Complex shapes, e.g., rising and then falling, 
are formed by combining multiple \ssgs 
using one or more \emph{operators}. 
One can search for multiple patterns in a sequence (concat, \colorconcat) or
matching the same sub-region of the trendline (and, \colorand), 
or one of many patterns matching a sub-region (or, \coloror), described later.

As an example,``rising from $x$=$2$ to $x$=$5$ and then falling'' 
can be translated into a \sq  [\loc{x.s=2,x.e=5}, \pat{p=up}] \colorconcat[\pat{p=down}] 
consisting of two \ssgs separated by a \colorconcat\ operator. 
The first \ssg captures ``rising from $x=2$ to $x=5$''; the second expresses a  ``falling" pattern. 
Since the second must ``follow'' the first, 
the two \ssgs are combined  using the \concat operator, 
denoted by \colorconcat. We now describe the shape primitives and operators that constitute \sq algebra. Table~\ref{tab:algebra} lists these primitives and operators.

\eat{
\begin{table}[t]
	\begin{center}
		\papertext{\vspace{-10pt}}
		\caption {\smallcaption{Primitives and Operators in  ShapeQuery}}
		\label{tab:algebra}
		\vspace{-10pt}
		\resizebox{0.75\columnwidth}{!}{%
			\begin{tabular}{|l l l |} 
				\hline
				Symbol & Name & Type \\ [0.5ex] 
				\hline
				\xs & {\sf START X VALUE} & Location Sub-Primitive  \\
				\ys & {\sf START Y VALUE} & Location Sub-Primitive  \\
				\xe & {\sf END X VALUE} & Location Sub-Primitive \\
				\ye & {\sf END Y VALUE} & Location Sub-Primitive  \\ 
				 \vs & \sketch & Location Sub-Primitive  \\	
				 \ld & \iterator & Location Sub-Primitive  \\ \hline
				\p  & \pattern & Primitive  \\
				\pd & \POSITION & Pattern Sub-Primitive  \\ \hline
	
				\m  & \modifier 	& Primitive   \\
				\mgt  & {\sf MORE}	& Modifier value   \\
				\mgteq & {\sf ATLEAST 2X} & Modifier value   \\
				\meq  & SIMILAR 	& Modifier value   \\
				\colorconcat  & \concat & Operator   \\
				\colorand & \AND & Operator  \\
				\coloror & \OR & Operator   \\
				$!$ & \opposite & Operator  \\
				\hline
			\end{tabular}%
		}
	\end{center}
	\vspace{-15pt}
\end{table}
}

\vspace{-5pt}
\subsection{Shape Primitives and Operators}
A \ssg is described using two high level primitives: \location and  \pattern.
\ssr allows users to skip one or more of these primitives in their query. The \location values can be skipped in order to match the 
\pattern anywhere in the trendline. 
Similarly, users can input the exact trendline to match, 
or the endpoints of the \ssgs to match without specifying the \pattern. 
We describe each of these supported primitives.

\stitle{Specifying \location}.
\location defines the endpoints of the sub-region of the trendline 
between which a pattern is matched: starting X/Y coordinate (\loc{x.s}/\loc{y.s}), 
ending X/Y coordinate (\loc{x.e}/\loc{y.e}).
For example, [\loc{x.s=2,x.e=10, y.s=10,y.e=100}] is a simple \sq to find trendlines 
whose trend between $x$=$2$ to $x$=$10$ is similar to the line segment from ($2,10$) to ($10,100$). Users can also draw a sketch to find trendlines
similar to the sketch, a functionality supported in 
other tools alluded to in the introduction~\cite{zenvisagevldb,timesearcher,googlecorrelate}.  
\ssr translates the pixel values of the 
user-drawn sketch to the domain values 
of the X and Y attributes, 
and adds the transformed vector of $(x,y)$ values as a vector $v$ in the \sq. 
As an example, the \sq [\loc{v=(2:10,3:14,...,10:100)}] 
finds trendlines that have precisely similar values 
to $v$ using a distance measure, e.g., Euclidean distance, or dynamic time warping~\cite{dtw}.


\stitle{Specifying \pattern}. 
\pattern defines a trend or a semantic feature in a sub-region of the trendline. 
A number of basic semantic patterns,  commonly used for characterizing trendlines, are supported,
such as {\em up}, {\em down}, {\em flat}, or the slope ($\theta$) in degrees. 
For example [\pat{p=up}] finds trendlines that are increasing, 
[\pat{p=45}] finds trendlines that are increasing with a slope of about $45^{\circ}$, and [\loc{x.s=2,x.e=10}, \pat{p=up}] 
finds trendlines that are increasing from $x=2$ to $10$. Finally, one can use 
\pat{p=*} to match any pattern and \pat{p=empty} to ensure that there are no points over the sub-region. 


\begin{table*}
	\vspace{-25pt}
	\parbox{.60\linewidth}{
		\centering
		\papertext{\vspace{-5pt}}
		\caption{\smallcaption{\rev{Examples of ShapeQueries}}}
		\vspace{-10pt}
		\resizebox{\linewidth}{!}{%
			\label{tab:examples}
			\begin{tabular}{ |p{8cm}|p{5cm}| } 
				\hline
				\textbf{Pattern} & \textbf{ShapeQuery} \\ \hline
				Increasing from 2 to 5 and then decreasing & [\pup, {\color{brown} x.s=2, x.e=10}]\colorconcat
				[\pdown] \\ \hline
				Decreasing or increasing anywhere & [{\color{olive} p = *}]\colorconcat(\pup \coloror \pdown) \colorconcat [{\color{olive} p = *}] \\ \hline
				Increasing at 45, decreasing
				at 60  and then becomes flat
				& [{\color{olive} p = 45}]\colorconcat [{\color{olive} p = -60}]\colorconcat [{\color{olive} p = flat}]  \\ \hline
				Decreasing over a width of
				3 points:
				& [{\color{brown} x.s=., x.e=.+3}, \pdown] \\ \hline
				Increasing at least once and at most 5 & [\pup, {\color{purple} q={1,5}}] \\ \hline
				W shaped pattern & [{\color{olive}
					p=-45}]\colorconcat [{\color{olive} p = 60}]\colorconcat[{\color{olive} p=-45}]\colorconcat[{\color{olive} p = 60}]   \\ \hline
				Specific sketch & [{\color{brown} v = (2:10,3:14,...,10:100)}]  \\ \hline
				Shape whose trend is increasing relatives its own trend before some point in the past (e.g, inverted bell shaped) & [\pdown]\colorconcat[{\color{olive} p > \$$-$.p}]\\ \hline
			\end{tabular}
		}
	}
	\hfill
	\parbox{.22\linewidth}{
		\centering
		\papertext{\vspace{-5pt}}
		\caption{\smallcaption{Pattern Scores}}
		\vspace{-10pt}
		\resizebox{\linewidth}{!}{%
			\label{tab:scoresgmt}
			\begin{tabular}{ |p{0.74cm}|c| } 
				\hline
				P & Score  \\  \hline
				{\em up} & $\frac{2 \cdot tan^{-1}(slope)}{\pi}$  \\ 
				{\em down} & $-\frac{2 \cdot tan^{-1}(slope)}{\pi}$  \\ 
				{\em flat} & $(1.0 - \lVert \tfrac{4 \cdot tan^{-1}(slope)}{\pi} \rVert)$  \\
				$\theta=x$ & $(1.0 - \lVert \tfrac{2 \cdot tan^{-1}(slope - x))}{(\pi-\rVert tan^{-1}(x \rVert)} \rVert)$ \\
				$*$ & 1 \\
				{\em empty} & -1 \\
				$v$ & $L_2$ norm (configurable) \\
				\hline
			\end{tabular}
		}
	}
	\hfill
	\parbox{.16\linewidth}{
		\centering
		\caption{\smallcaption{Operator Scores}}
		\vspace{-10pt}
		\label{tab:scoreopt}
		\resizebox{\linewidth}{!}{%
			\begin{tabular}{ |c|c| } 
				\hline
				O & Score \\  \hline
				\colorconcat & $ \sum_{i=1}^{k}score_i/k$ \\ 
				\colorand & $min(score_1, ...,score_k)$ \\
				\coloror & $max(score_1, ...,score_k)$\\
				\hline
			\end{tabular}%
		}
	}
	\vspace{-10pt}
\end{table*}

\stitle{Combining \paterns.}
\sq supports three operators to combine \ssgs: 
\begin{denselist}
\eat{\item \match ([\hspace{0.1cm}]) is a unary operator that takes as input a \ssg, 
executes it over one or more sub-regions (visual segments) of the trendline.  As a rule, every \ssg is bounded to the MATCH operator.}

\item \concat (\colorconcat) specifies a sequence of two or more \ssgs. For example, using \pat{[p=up]}\colorconcat\pat{[p=down]}  one can search for genes that are first rising, and then falling. Note that {\colorconcat} is one of the most frequently used operations, and we sometimes omit {\colorconcat} between \ssgs, e.g, \pat{[p=up]}\pat{[p=down]}, to make it succinct to describe.

\item \AND (\colorand)  simultaneously matches multiple patterns in the same sub-region of the trendline. Unlike \concat, all of the patterns must be present in the same sub-region. For example, one can look for genes whose expression values rise twice but do not fall more than once within the same sub-region.

\item \OR (\coloror)  searches for one among many patterns in the same sub-region of the trendline, picking the one that matches the sub-region best. For example, one can search for genes whose expressions are either {\em up}- or {\em down}-regulated.
 
\end{denselist}
Note that when the same operator is specified consecutively, \ssr fuses them into one, hence all operators can take two or more operands. For example, [\pat{p=up}] \colorconcat [\pat{p=down}] \colorconcat  [\pat{p=down}] is parsed as  a single {\colorconcat} operation with three operands [\pat{p=up}], [\pat{p=down}], and [\pat{p=down}]. 

Multiple operations are often used in a given \sq.  \ssr follows left to right precedence order for execution of the  operations. However, sub-expressions can be nested using parentheses $()$ to specify precedence as in mathematical expressions.  In Figure~\ref{fig:ast}, we depict how \ssr parses a complex \sq[{\color{olive} p=up}]\colorconcat[{\color{olive} p=down}] 
\colorconcat(([{\color{olive} p=up}]\colorconcat[{\color{olive} p=down}])\coloror[{\color{olive}p=flat}])
 into an Abstract Syntax Tree (AST) representation.

\emph{Comparing Patterns.} In some cases, one may want to compare the pattern in a \ssg with the preceding or succeeding \ssgs. To support such use cases, \ssr (i) allows a \ssg to refer to the previous or the next \ssg using $\$+$ or $\$-$ respectively, 
and (ii) compare patterns between the current and referred \ssg using operations $>$, $<$, or $=$. For example, astronomers can issue a \sq [\pat{p=up}]\colorconcat[{\color{olive} p < \$$-$.p}] with $x$=time and $y$=luminosity (brightness)  to search for celestial objects that were initially moving rapidly 
towards earth, but after some point either slowed down  
or started moving away. The second \ssg [{\color{olive} p < \$$-$.p}]
ensures that the slope of brightness over time is less than that 
in the previous sub-region [\pat{p=up}].

Similarly, one can set {\color{olive} p <$\frac{1}{2}$\$$-$.p} to ensure the slope of second sub-region is  $\leq \frac{1}{2}$ of the first. To avoid ambiguity in position reference and for efficient execution, \ssr restricts \$-based references to a simple \concat operation, i.e., across a sequence of patterns at the same level of nesting.

\eat{
\begin{figure}
	\papertext{\vspace{-5pt}}
	\centerline {
		\hbox{\resizebox{7.5cm}{!}{\includegraphics{figs/sqtree.pdf}}}}
	\vspace{-10.5pt}
	\caption{\smallcaption{The ShapeQuery Tree}}
	\vspace{-8.5pt}
	\label{fig:segtree-intuition}
	\vspace{-1.5pt}
\end{figure}
}

\eat{
\stitle{Grouping and Nesting.}
\ssr follows left to right precedence order for execution of operations. However, sub-expressions can be grouped using parentheses $()$ to specify
precedence as in mathematical expressions.  In Figure~\ref{fig:ast}, we depict how \ssr parses a complex \sq [\pat{p=up}]\colorconcat 
[\pat{p=down}] \colorconcat(([\pat{p=up}])\colorconcat[\pat{p=down}]) \coloror[\pat{p=flat}]
into an Abstract Syntax Tree (AST) representation. \tar{re: part of parsing: this is only a tiny bit of parsing info I wanted to give here for algorithms to make sense later.}
}

\stitle{Expressing complex patterns.} The aforementioned basic primitives and operators are powerful enough to express more complex \ssr use-cases. We discuss three such complex patterns below, along with shortcuts for their easy specification.

\emph{1. Searching shapes of specific width.} In some cases, users want to find specific shapes irrespective of their start location. For example, one may want to  search for cities  with maximum rise in temperature  over a width of $3$ months.
To express such queries, \ssr supports the \iterator (.), e.g., [\loc{x.s=.,x.e=x.s+3},\pat{p=up}] that iterates over all points in the trendline, setting each point as the start $x$ position, with the $x$ end position set to 3 units ahead.  Internally, for a trendline of length $n$, this query can be rewritten as an \OR operation over  $(n-3+1)$ \ssgs, where, for the $i$th \ssg, \loc{x.s=i} and \loc{x.e=i+3}. 

\emph{2. Quantifiers.} One can search for trendlines where a pattern occurs a specific number of times using quantifiers, denoted by \md{q}. For example, [\pat{p=up},\md{q={1,2}}] can be used to search for trendlines where there is an increasing pattern at least once and at most twice. Quantifiers can be internally rewritten using an \OR of one or more \concat operations. For example, the above query is rewritten as ([\pat{p=*}] \colorconcat[\pat{p=up}]\colorconcat[\pat{p=*}])\coloror ([\pat{p=*}]\colorconcat[\pat{p=up}]\colorconcat[\pat{p=*}] \colorconcat[\pat{p=up}]\colorconcat [\pat{p=*}]).

\emph{3. Nesting} A combination of patterns can be constrained to be within a specific sub-region by specifying them as a value of the \pattern primitive. For example,  to search for stocks that increased anytime between February to October, we can use nesting as follows: [\loc{x.s=2,x.e=10}, \pat{p}=([\pat{p=*}][\pat{p=up}][\pat{p=*}])]. This can be rewritten using \concat operations as follows: [\loc{x.s=2},\pat{p=*}]\colorconcat
[\pat{p=up}]\colorconcat[\pat{p=*}]\colorconcat[\loc{x.s=10},\pat{p=*}].

\emph{4. Scale invariant matching.} \agptechreport{One}\agppapertext{Finally, one} can  automatically  search for shapes at varying granularity of x-scales and degree of smoothing. For example, for [\pat{p=*}]\colorconcat[\pat{p=up}] \colorconcat[\pat{p=down}]\colorconcat[\pat{p=*}], \ssr searches for [\pat{p=up}] and then [\pat{p=down}] at all possible scales and selects the one that leads to the best match. To do so, \ssr uses efficient algorithms that we describe in the subsequent sections.

As \ssr evolves, it may support additional shortcuts to simplify the writing of frequently used complex patterns.However, all of the complex patterns as well as the shortcuts can be expressed using basic primitives and operations for their execution. Thus, we omit further discussion of complex patterns and limit ourselves to the semantics and efficient scoring of basic primitives and operators.

\eat{
\begin{figure}
\centering
\resizebox{0.40\columnwidth}{!}{%
\begin{tikzpicture}[node distance=2cm]
\node (A) at (1.2, 3) {\colorconcat};
\node (B) at (0, 2) {[\pup]};
\node (C) at (1.2, 2) {[\pdown]};
\node (D) at (2.5, 2) {\coloror};
\node (E) at (1.2, 1) {\colorconcat};
\node (F) at (3, 1) {[\pup]};
\node (G) at (0.5, 0) {[\pup]};
\node (H) at (1.9, 0) {[\pdown]};
\draw[thick,->] (A) -- (B);
\draw[thick,->] (A) -- (C);
\draw[thick,->] (A) -- (D);
\draw[thick,->] (D) -- (E);
\draw[thick,->] (D) -- (F);
\draw[thick,->] (E) -- (G);
\draw[thick,->] (E) -- (H);
\end{tikzpicture}%
}
\caption{Abstract tree representation of \sq [{\color{olive} p=up}]\colorconcat 
	[{\color{olive}p=down}]\colorconcat(([{\color{olive} p=up}]\colorconcat[{\color{olive} p=down}])\coloror[{\color{olive} p=flat}])
}
\label{fig:ast}
\end{figure}
}

\vspace{-5pt}
\subsection{Formal semantics of  \sq}
We now formally define the semantics of \sq.

Given three dataset attributes $x$, $y$, and $z$, \
\ssr first generates a collection of trendlines $V$, 
one for each unique value of the $z$ attribute. 
Each trendline is a sequence  of ($x$, $y$) values ordered by $x$. 
A \sq $Q$ operates on one trendline,  $V_i$, at a time, 
and returns a real number, called {\em score}, between $-1$ to $+1$, i.e.,  
$Q: V_i \rightarrow score$; $score \in [-1,1]$. 
The value of $score$ describes how closely $V_i$ matches $Q$, 
with $+1$ the best possible match, and $-1$ the worst.

The \sq Q operates on $V_i$ 
with the help of \ssgs ($S_1, S_2,\ldots,S_n$) and 
operators ($O_1, O_2,\ldots,O_m$).  
Each \ssg, $S_i$ operates on $V_i^{p,q}$, 
a sub-region of $V_i$ starting at $p=x.s$ and 
ending at $q=x.e$ and 
returns a $score_i \in [-1,1]$ 
using scoring functions we describe subsequently. 
A common subclass of \sqs are \emph{fuzzy \sqs}. A fuzzy \sq is a sequence of \ssgs where there is at least one \ssg with missing or multiple possible values for $x.s$ or $x.e$. Thus, for fuzzy \sqs, we try all possible values of $p$ and $q$, selecting the sub-region that leads to the best score. 
One or more \ssgs are combined using operators such as \colorconcat, \colorand, \coloror. Formally, an operator $O_i$ takes as input the scores $score_1, score_2, \ldots, score_n$  from its $n$ input \ssgs and outputs another $score_i$ 
using scoring functions that capture the behavior of the operators. When combined via \AND or \OR operators, \ssgs may operate on overlapping sub-regions $V_i^{p,q}$, however, for \concat, the sub-regions must not overlap since \concat specifies a sequence of patterns. Next, we describe our scoring methodology.

\eat{
Overall, we represent \sq (Q) with a context-free grammar ($CFG$) using the following set of rules:

\begin{denselist}
\item     $ Q \rightarrow S | Q \otimes S | Q \odot S | Q\oplus S | !Q | (Q) $

\item $  S \rightarrow  [(L,)^+ (P,)^+ (M,)^+] $ 

\item $ L \rightarrow (x.s=num,)^+ (x.e=num,)^+ (y.s=num,)^+ (y.e=num,)^+ | (x.s=., x.e=.+num) |  v = (num:num,)^* $ 

\item $ P \rightarrow  (p=up|down|flat|num|\$num|udp|S) $

\item $ M \rightarrow  q = \{num^+,num^+\}| 
\end{denselist}
}

\subsection{Scoring Methodology}
\label{sec:scoring}

For supporting interactive response times, 
\ssr needs to \emph{efficiently} and \emph{effectively} compute the match between  a \sq $Q$ and a trendline $V_i$.

To satisfy both efficiency and effectiveness, 
\ssr approximates each sub-region with a line, 
using the slope to quantify how closely it captures any given \ssg. 
The line-based quantification is robust to noise or minor fluctuations, as is 
often intended in \sqs. At the same time, lines are extremely fast to compute,  requiring only a single pass on the data.  
As we explain shortly, lines over larger sub-regions 
can be quickly inferred from lines over over smaller ones, 
without additional passes. \rev{As the complexity of a pattern increases, the number of lines required to approximate it also increases. However, even for complex shapes, a small number of line segments is sufficient. Our study of patterns (e.g.,  double-bottom, triple-top) in finance~\cite{ge1998pattern} as well as mturk queries reveal that the maximum number of lines is usually small (typically less than $6$). 
}

\eat{
  \begin{table}
	\vspace{-25pt}
	\parbox{.52\linewidth}{
		\centering
		\papertext{\vspace{-5pt}}
		\caption{\smallcaption{Pattern Scores}}
		\vspace{-10pt}
		\resizebox{0.55\columnwidth}{!}{%
			\label{tab:scoresgmt}
			\begin{tabular}{ |p{0.74cm}|c| } 
				\hline
				P & Score  \\  \hline
				{\em up} & $\frac{2 \cdot tan^{-1}(slope)}{\pi}$  \\ 
				{\em down} & $-\frac{2 \cdot tan^{-1}(slope)}{\pi}$  \\ 
				{\em flat} & $(1.0 - \lVert \tfrac{4 \cdot tan^{-1}(slope)}{\pi} \rVert)$  \\
				$\theta=x$ & $(1.0 - \lVert \tfrac{2 \cdot tan^{-1}(slope - x))}{(\pi-\rVert tan^{-1}(x \rVert)} \rVert)$ \\
				$*$ & 1 \\
				{\em empty} & -1 \\
				$v$ & $L_2$ norm (configurable) \\
				\hline
			\end{tabular}
		}
	}
	\hfill
	\parbox{.44\linewidth}{
		\centering
		\caption{\smallcaption{Operator Scores}}
		\vspace{-10pt}
			\label{tab:scoreopt}
		\resizebox{0.44\columnwidth}{!}{%
			\begin{tabular}{ |c|c| } 
				\hline
				O & Score \\  \hline
				\colorconcat & $ \sum_{i=1}^{k}score_i/k$ \\ 
				\colorand & $min(score_1, ...,score_k)$ \\
				\coloror & $max(score_1, ...,score_k)$\\
				\hline
			\end{tabular}%
		}
	}
	\vspace{-10pt}
\end{table}
}

As depicted in Table~\ref{tab:scoresgmt}, \ssr uses different scoring functions 
for each pattern primitive that transforms the slope 
to a value in $[-1,1]$ using a $tan^{-1}$ function. 
For example, for an {\em up} pattern, the function returns 
a score between $[0,1]$ for all trendlines 
with slope from $0^\circ$ to $90^\circ$, 
a score of $[-1,0]$ for slopes $< 0^\circ$ (opposite of {\em up}). 
Moreover, a change in trend from  $10^\circ$ to $30^\circ$ 
is visually more noticeable 
than from $60^\circ$ to $80^\circ$, 
thus we capture this behavior using $tan^{-1}$ 
where the rate of increase in output decreases 
as the value of slope increases. Finally, we apply normalization, such as multiplying by $2/\pi$, to re-scale the output of $tan^{-1}$ between $-1$ and $1$.  Thus, depending on the specified pattern primitive, \ssr uses the corresponding scoring function to compute the score for that \ssg. For a given \ssg, if the location constraints are not met, we assign a score of -1, and ignore the rest of the primitives.

We state the following observation regarding the scoring of a single \ssg.

\stitle{Observation 2.1.} \emph{The scoring of a \ssg, as part of a \sq $Q$ without comparisons, on a sub-region $L$ can be done using the slope of the corresponding single line segment and the $x.s$, $x.e$, $y.s$, and $y.e$ values of the sub-region, independent of other sub-regions.}

\revtr{For a shape input as a sketch, users sometimes intend to
perform precise matching. For such ShapeSegments we compute the score using L2 norm (Euclidean distance) between
the drawn sketch and the trendline without fitting a
line segment. The L2 norm can vary from 0 to $\infty$; therefore, we normalize the distance within [1, -1] using using Max-Min normalization~\cite{minmax}. In addition, \ssr allows users to use sketch for fuzzy matching where \ssr fits a minimum number of lines to the sketch given an error threshold (adjustable via a slider), and automatically constructs a CONCAT operation of  \ssgs, with one \ssg for each line with the pattern corresponding the slope of the line. We provide more details on fuzzy matching using sketch in Appendix~\ref{sec:sketchapprox}.
}

\techreport{In addition to providing domain-specific pattern types
(UDPs), ShapeSearch also allows users to override the default scoring methodology by letting them define their own
scoring functions. For seamless integration, user-defined
scoring functions must take a sun-region as input, and
output a score within [−1, 1].}

For two contiguous \ssgs compared using \$-references, \ssr returns a single score as if they were one single \ssg evaluated over their combined sub-region. Internally, \ssr evaluates each of \ssgs over their corresponding sub-region independently and combines scores across the \concat appropriately. The score of the \ssg that uses that \$ reference is set to $+1$ if the constraint is satisfied, otherwise it is set to $-1$. For ease of explanation, we refer them as a single \ssg for the rest of the paper.


 The scores across \ssgs are combined using the scoring functions for the operations. Note that, in general, as depicted in Figure~\ref{fig:ast}, the operands of an operator can be sub-expressions involving other operators. Nevertheless, as depicted in Table~\ref{tab:scoreopt}, the scoring functions for operators are more straightforward as they directly capture the semantic behavior of the operators.
For instance, \concat matches a sequence of patterns, 
therefore, the scoring function takes average of the scores of its operands 
to give equal weightage to each operand. \AND matches multiple patterns over the same sub-region, so to avoid any \ssg not having a good match, 
we take the minimum of all scores across its operands. On the other hand, \OR picks the best among all matches, so it takes the maximum across all scores.
From these definitions, we state the following observations:

\stitle{Observation 2.2.} \emph{
	The scoring of AND or OR operations with $k$ operands on a sub-region $L$ can be done by scoring each of the $k$ operands independently on the sub-region $L$.}

\stitle{Observation 2.3.} \emph{
	The scoring of {\concat} with $k$ operands on sub-region $L$ can be done by dividing sub-region $L$ into all possible sequences of $k$ sub-regions, followed by scoring operand $i$ on sub-region $i$.}

Note that the scoring of an operand can be done independent of others. We used the term \emph{segmentation} to refer to a division of a sub-region into $L$ sub-regions.


\stitle{Ensuring goodness of fit.} It is possible that a line poorly approximates a given segment of the trendline. 
Therefore, we use a configurable (via a slider) threshold parameter to suggest how much error can be tolerated. 
For measuring the goodness of fit, we compute the standard $R^2$ error~\cite{r2wiki}, also called coefficient of determination, of the line, between $0$ to $1$, 
with higher values indicating lower errors and better fit. 
\ssr gives a score of $-1$ to a \ssg for a given sub-region if $R^2$ is less than the threshold.

\stitle{Overall algorithm.} Algorithm~\ref{algo:scoringalgo} outlines the steps for scoring a \sq. At the start, the algorithm takes the entire trendline $V_i$ as $L$, the  Abstract Tree Representation (AST) of \sq as $Q$, and the list of scoring functions  $ScrFunc$ as in Tables~\ref{tab:scoresgmt} and ~\ref{tab:scoreopt} as inputs. If the root node of the \sq tree is a \ssg, \ssr directly computes the score of the \ssg on the sub-region using scoring functions after checking the location and goodness of fit constraints (lines $2$-$9$). If the root node is {\colorand} or {\coloror}, \ssr invokes each of the operands (i.e., child sub-trees) to compute their scores on the sub-region independently, combining the scores as per their scoring functions (lines $14$-$18$). However, if the root node is a \concat with $k$ operands, i.e., child sub-trees, \ssr segments $L$ into all possible $k$ sub-regions: $L_1, L_2, ..., L_k$, and then for each segmentation, invokes the ith operand on ith segment (lines $20$-$30$). 
Finally, the maximum score across all segmentations is output.

\begin{algorithm}
	\caption{\sq Scoring}
	\label{algo:scoringalgo}
	\small
	\begin{flushleft}
		\textbf{Input:} $L$: a sub-region of trendline, $Q$: a \sq sub-expression, $ScrFunc$: scoring functions from Tables~\ref{tab:scoresgmt} and ~\ref{tab:scoreopt} \\
		\textbf{Output:} score
	\end{flushleft}
	\begin{algorithmic}[1]
		\Procedure{ExecShapeQuery}{$L$, $Q$, $ScrFunc$} 
		\If{$Q.root$ is a \ssg }
		\State $hasValidLoc$ <-	CheckLocationConstraints($L$, $Q$)
		\State $hasValidLineFit$ <-	CheckGoodnessofFit($L$)
		\If{($hasValidLoc$ \&\& $hasValidLineFit$)  is False}
		\State return -1;
		\EndIf
		\State return $ScrFunc(L,operator,Q.root)$
		\EndIf
		\State $operator$ $\gets$ $Q.root.operator$
		\State $operands$ $\gets$  $operator.children$
		\State $k  = operands$.size
		\State $operandscores$ $\gets$ []  
		\If{$operator$ $\in$ $\{$\colorand,\coloror$\}$}
		\For{each $child$ in $operands$}
		\State	$operandscores$.append(\textsc{ExecShapeQuery}($L$,$child$))
		\EndFor
		\State return $ScrFunc(operator,operandscores)$
		\EndIf
		\If{$operator$ $\in$ $\{$\colorconcat$\}$}
		\State $candscores$ = []
		\For{each segmentation $\{L_1, L_2, ..., L_k\}$ of $L$}
		\State $sgtscores$ $\gets$ []  
		\For{each $child$ in $operands$}
		\State	$sgtscores$.append(\textsc{ExecShapeQuery}($L_i$,$child$))
		\EndFor
		\State $sgtscores$$\gets$ ScoreComparators($sgtscores$,$Q.root$)

		\State $candscores$.append($ScrFunc$($operator$,$sgtscores$))
		\EndFor
		\State	return $max(candscores)$
		\EndIf
		
		\EndProcedure
	\end{algorithmic}
\end{algorithm}


\section {Executing Fuzzy ShapeQueries}
\label{sec:fuzzychallenge}

The most interesting and powerful feature of \ssr is its capability for ``fuzzy'' matching,  allowing users to search for patterns 
without specifying exact locations, e.g., 
increasing followed by decreasing. Recall that a a \emph{fuzzy \sq is one
with atleast one \ssg with multiple possible values for $x.s$ or $x.e$}.

In the absence of exact location values for a \concat, \ssr has to exhaustively score all possible segmentations to find the one with the best $score$ (line 22-29 in Algorithm 1).  This becomes prohibitively expensive as the number of points in the trendline increases. 
For example, a fuzzy \sq [\pat{p=up}]\colorconcat[\pat{p=down}]\colorconcat[\pat{p=up}] 
on a trendline with $100$ points 
can result in $10^4$ possible segmentations for finding three segments that lead to the best score.  
More generally, for a \concat with $k$ operands, 
the exhaustive approach creates $n^{(k-1)}$ segmentations, 
where $n$ is the number of points in the trendline.  

We state this problem formally:
\begin{problem}[Fuzzy CONCAT Scoring]	Given a CONCAT operation with $k$ operands	 and a sub-region $L$ of the trendline with $n$ points, find the segmentation with $k$ subregions where the score of CONCAT is maximum.
\end{problem}
\vspace{-10pt}

\subsection{The Dynamic Programming Algorithm} 
\label{subsec:dp}

We first show that we can substantially reduce the number of segmentations for a \concat operation on a sequence of \ssgs by reusing the scores from \concat operations over sub-sequences of \ssgs. We, then, show that this extends to the case when one or more operands of \concat are AND or OR expressions, but none of the operands internally involve nested {\concat}s. Finally, we show how we can reuse computations when an  AND or OR operand internally has a nested {\concat} or when an operand is a nested \concat. We start with the simplest case of a \concat on a sequence of \ssgs.

From Observation 2.3, it can be seen that for the \concat operation itself, the scoring of the $jth$ operand on $jth$ sub-region does not depend on the scoring of the first $j-1$ operands on the first $j-1$ sub-regions.
Thus, we can find the optimal segmentation of the first $j-1$ operands over all smaller sub-regions and combine them with the scores of $jth$ operand on the remaining part of the sub-region to find the optimal segmentation.

Suppose the optimal segmentation of [\pat{p=up}]\colorconcat[\pat{p=down}] \colorconcat[\pat{p=flat}] over sub-region \loc{x=1} to  \loc{x.e=100} is when [\pat{p=up}] is scored over the sub-region  \loc{x.s=1} to  \loc{x.e=45}, [\pat{p=down}] over \loc{x.s=46} to  \loc{x.e=60}, and  [\pat{p=flat}] over  \loc{x.s=61} to \loc{x.e=100}. Then, for another CONCAT operation involving a sub-sequence [\pat{p=up}]\colorconcat[\pat{p=down}] over the sub-region  \loc{x.s=1} to  \loc{x.e=60}, the optimal segmentation should have the same sub-regions for [\pat{p=up}] and [\pat{p=down}] as in the previous CONCAT. This is because the scoring of [\pat{p=flat}] from \loc{x.s=61} to  \loc{x.e=100} does not affect the scoring of [\pat{p=up}]\colorconcat[\pat{p=down}] over \loc{x.s=1} to  \loc{x.e=60}. 

We use this idea to develop a faster dynamic programming algorithm (DP) for scoring CONCAT operations over \ssgs. Formally, let  $OPT(1,t,(1:j-1))$ be the best score corresponding to the optimal segmentation over the sub-region between $x=1$ to $x=t$ for first $j-1$ operands, and $SC(t+1,i,j)$ be the score of $jth$ operand over the sub-region between $x=t+1$ and $x=i$. Then, the optimal segmentation $OPT(1,i,(1:j))$ for first $j$ operands over $x=1$ and $x=i$ can be computed using the following recursion:

\noindent
{	$OPT(1,i,(1,j))$
	$=\underset{t}{MAX}\{\frac{(j-1)\times OPT(1,t,(1:j-1))+SC(t+1,i,j))}{j}\}$
}

\agptechreport{
As base cases, we set $OPT(m,m+1,(j:j)) = SC(m,m+1,j)$.
}

Using the above recurrence, we develop a DP algorithm that reuses  intermediate results by memoizing $OPT(1,i,(1,j))$ in a $2D$ array of size $n \times k$  and $SC(t+1,i,j)$ in $3D$ array of size $n\times n \times k$. The DP algorithm considers $O(n^2k)$ segmentations to find the optimal score.

\begin{theorem} 
 	Finding the best segmentation for a CONCAT operation on \ssgs arguments can be done in $O(n^2k)$ using Dynamic Programming.
 \end{theorem}
\vspace{-8pt}

\stitle{ AND/OR operands with no nested CONCAT.} Let the $i$th operand of the \concat at the root be an AND/OR expression with no nested CONCAT. From Observation 3.2, and lines 17-22 in Algorithm 1, we can score operand $i$ on sub-region $i$ without any segmentation. Thus, the  above theorem is also valid when an operand in the \concat operation is an AND/OR expression with no CONCAT operations internally.

\stitle{ AND/OR operand with a nested CONCAT or directly nested CONCATs.} If the $i$th operand of the \concat at the root consists of a nested \concat (either under an AND or OR expression or directly), the $i$th sub-region needs to undergo further segmentation to find the optimal score for the nested \concat. For example, for scoring the \sq in Figure~\ref{fig:ast}, the DP algorithm is first invoked for the \concat at the root node. The third operand for the root \concat is an OR which consists of another \concat and [\pat{p=flat}] as its operands. Therefore, for every candidate sub-region for the third operand, another DP algorithm is invoked for the nested \concat. However, this is not a problem since we can reuse the score of \ssgs across invocations. For example, in Figure~\ref{fig:ast}, \concat operations essentially involve scoring of \ssgs [\pat{p=down}], [\pat{p=up}] over all possible sub-regions. We, thus, score  each  \ssg only once for a given sub-region, and reuse it across multiple invocations of the DP algorithm for each \concat. Moreover, the DP recursion involves $SC(t+1,i,j)$ for computing the cost of the operand (i.e., sub-expression) $j$ from sub-region $t+1$ to $i$, which again can be shared across repeated invocations of the same $t,i,j$. \smallskip
 
\noindent
Unfortunately, even though the DP algorithm is orders of magnitude faster than the exhaustive approach, we note that for trendlines with large number of points, even a \sq with a single \concat operation can be slow, because of its quadratic runtime. As we will see  in  Section~\ref{sec:exp}, the DP algorithm takes 10s of seconds even for \sqs with 3 or 4 \ssgs over trendlines with a few hundreds of points. We, next, discuss optimizations to further decrease the runtime of CONCAT operation on \ssgs.

\eat{
\begin{figure}
	\papertext{\vspace{-5pt}}
	\centerline {
		\hbox{\resizebox{7.5cm}{!}{\includegraphics{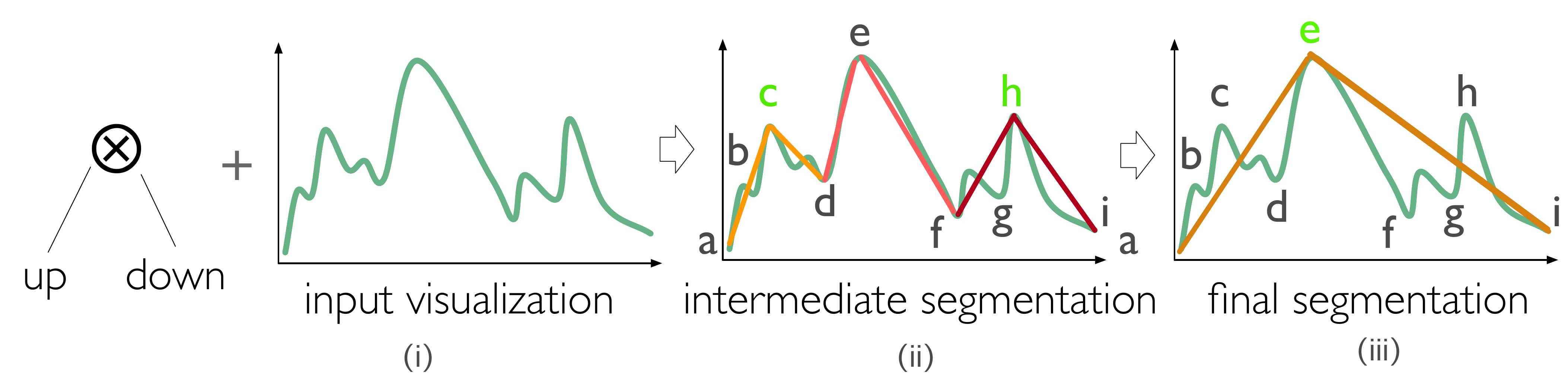}}}}
	\vspace{-10.5pt}
	\caption{\smallcaption{Intuition behind Pattern-aware segmentation }}
	\vspace{-8.5pt}
	\label{fig:segtree-intuition}
	\vspace{-1.5pt}
\end{figure}
}

\subsection{A Pattern-Aware Bottom-up Approach}
\label{subsec:segmenttree}

\eat{
\begin{figure}
	\centering
	\resizebox{0.5\columnwidth}{!}{%
		\begin{tikzpicture}[node distance=2cm]
		\draw [rounded corners = 100mm,cyan, thick]
		(0,0.2) -- (0.5,0.6)
		(0.5,0.6) -- (1,1)
    	(1,1) -- (1.5,0.6)
   	 	(1.5,0.6) -- (2.0,0.2)
  	 	(2,0.2) -- (2.5,0.70)
   	 	(2.5,0.70) -- (3,0.45)
		(3,0.45) -- (3.5,0.80)
		(3.5,.80) -- (4,0.5) ;
		\draw [<->, black] (0,1.2) to (1.99,1.2);
		\draw [<->, black] (2.01,1.2) to (4,1.2);
		
		\node[text width=0.2cm] at (1,1.35) {A};
		\node[text width=0.2cm] at (3,1.35) {B};

	    \draw [<->, black] (0,0) to (1,0);
		\draw [<->, black] (1,-0.2) to (2,-0.2);
		
		\node[text] at (0.5,-0.2) {R1};
		\node[text] at (1.5,-0.4) {R2};				
	   \end{tikzpicture}%
	}
	\caption{Abstract tree representation of \sq [{\color{olive} p=up}]\colorconcat 
		[{\color{olive} p=down}] 
		\colorconcat(([{\color{olive} p=up}]\colorconcat[{\color{olive} p=down}]) \coloror[{\color{olive} p=flat}])
	}
	\label{fig:ast}
\end{figure}
}

The DP-based optimal approach scores all possible sub-regions for each operand in the \concat operation. For instance, consider  a fuzzy \sq [\pat{p=up}]\colorconcat[\pat{p=down}]\colorconcat [\pat{p=flat}] and a trendline $L$ of $120$ points. Here, for operand [\pat{p=up}], the DP approach scores sub-regions of all possible sizes, starting from the smallest possible sub-region (\loc{x.s=1} to \loc{x.e=2})
to  (\loc{x.s=1} to \loc{x.e=116}). Note that a sub-region requires at least $2$ points to fit a line. 

\stitle{Greedy approach.} Two sub-regions that differ only in a few points tend to have similar scores. For instance, the scores of  [\pat{p=up}] over sub-regions (\loc{x.s=1} to \loc{x.e=15}) and (\loc{x.s=1} to \loc{x.e=16}) are likely to be similar.  
Therefore, an optimization over DP is to consider only those sub-regions for each \ssg that differ substantially in their sizes. 
For example, a greedy approach could be to start with sub-regions of equal size for each of the three \ssgs (i.e., 40 points each), and then greedily vary their sizes until we reach the maximum. One way of varying their sizes is to greedily extend one sub-region at a time, and proportionally shrink the others. For example, the next three configurations after starting with equal sizes could be: (60,30,30), (30,60,30), (30,30,60). We pick the best of these and then repeat the process. Clearly, this approach scores much fewer segmentations ($O(log(n^k))$), compared to $O(n^2)$ segmentations explored by the DP approach. However, as we show in our experiments (Section~\ref{sec:exp}), such an approach leads to extremely poor accuracy.

\stitle{Pattern-aware segmentation.}  The problem with the greedy approach is that it treats all points equally, and as possible candidates for endpoints of \ssgs. 
A better approach could be to select end points to be those where the slope (or pattern) changes drastically.
We first illustrate our intuition, and then describe an  algorithm that performs segmentation in a pattern-aware manner.
 
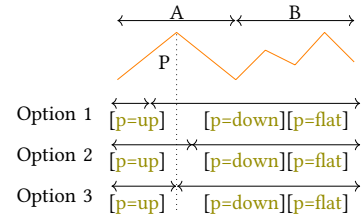
\begin{figure}
	\centering
	\resizebox{0.55\columnwidth}{!}{%
	\begin{tikzpicture}[node distance=2cm]
	\draw [rounded corners, orange]
	(0,0.2) -- (0.5,0.6)
	(0.5,0.6) -- (1,1)
	(1,1) -- (1.5,0.6)
	(1.5,0.6) -- (2.0,0.2)
	(2,0.2) -- (2.5,0.70)
	(2.5,0.70) -- (3,0.45)
	(3,0.45) -- (3.5,1)
	(3.5,1) -- (4,0.5) ;
	\draw [<->, black] (0,1.2) to (1.99,1.2);
	\draw [<->, black] (2.01,1.2) to (4,1.2);
	
	\node[text width=0.2cm] at (1,1.35) {A};
	\node[text width=0.2cm] at (3,1.35) {B};
	
	\draw [dotted, black] (1,-2.0) to (1,1);
	\node[text width=.2cm] at (0.8,0.5) {P};
	
	\node[text width=2cm] at (-0.7,-0.4) {Option 1};
	\draw [<->, black] (-0.1,-0.2) to (0.54,-0.2);
	\draw [<->, black] (0.56,-0.2) to (4.1,-0.2);
	\node[text width=0.1cm] at (-0.1,-0.5) {[\pup]};
	\node[text width=0.1cm] at (1.5,-0.5) {[\pdown][\pflat]};

	\node[text width=2cm] at (-0.7,-1.1) {Option 2};
	\draw [<->, black] (-0.1,-0.9) to (1.26,-0.9);
	\draw [<->, black] (1.26,-0.9) to (4.1,-0.9);
	\node[text width=0.1cm] at (-0.1,-1.2) {[\pup]};
	\node[text width=0.1cm] at (1.5,-1.2) {[\pdown][\pflat]};

	\node[text width=2cm] at (-0.7,-1.8) {Option 3};
	\draw [<->, black] (-0.1,-1.6) to (0.99,-1.6);
	\draw [<->, black] (1.01,-1.6) to (4.1,-1.6);
	\node[text width=0.1cm] at (-0.1,-1.9) {[\pup]};
	\node[text width=0.1cm] at (1.5,-1.9) {[\pdown][\pflat]};
	
	\end{tikzpicture}%
}
\vspace{-10pt}
\caption{Pattern-aware selection of LOPs} 
\label{fig:lop}
\vspace{-12pt}
\end{figure}

\stitle{Intuition.} As depicted in Figure~\ref{fig:lop}, consider two sub-regions $A$ on the left and $B$ on the right for the trendline $L$. Say the trendline in sub-region $A$ is  inverted V-shaped, i.e., increasing until a point $P$ and then decreasing. Now, for all possible  segmentations where [\pat{p=up}]'s sub-region lies completely in $A$, there are following possibilities for \loc{x.e} of [\pat{p=up}]: 
1) [\pat{p=up}]'s \loc{x.e} point is before $P$.
2) [\pat{p=up}]'s \loc{x.e} point is after $P$.
3) [\pat{p=up}]'s \loc{x.e} point is at $P$.

Since [\pat{p=down}] follows [\pat{p=up}], we can see that option 1 that sets [\pat{p=up}]'s  \loc{x.e} $<$ $P$ is less likely to be optimal as that will lead to scoring of a part of [\pat{p=down}] on an increasing trend. Similarly, \loc{x.e} $>$ $P$ is less optimal as that will lead to scoring of a part of [\pat{p=up}] on a decreasing trend. Thus, if we have to (greedily) select one point in sub-region $A$ for [\pat{p=up}]'s \loc{x.e}, $P$ is likely a better choice. We call such a point as \emph{locally optimal point} (LOP).

\stitle{A Bottom-up algorithm.} Based on the above intuition, we develop a much faster algorithm that uses the following assumption to reduce the number of segmentations.

\begin{assumption}[Closure]
	If a point is not locally optimal for any of the sub-expressions in the \concat operation (i.e., a \concat on a sub-sequence of the operands), it cannot be \loc{x.s} or \loc{x.e} of a \ssg in the optimal segmentation. 
\end{assumption}

That is, local optimality leads to global optimality. Due to this assumption, our proposed algorithm is approximate. However, our empirical results (Section~\ref{sec:exp}) show that despite this assumption, the accuracy of the algorithm is very close to that of DP, while taking orders of magnitude less time. 

\rev{ Algorithm~\ref{algo:segmentreealgo} outlines the steps for scoring a fuzzy \sq. At a high level, the  algorithm starts by dividing the trendline into smaller contiguous sub-regions (line 2)}. Next, it selects
locally optimal points (LOPs), defined next, over small sub-regions (line 12), followed by a bottom-up  merging step that uses LOPs over small sub-regions to find LOPs over larger sub-regions. 

\stitle{Selection of LOPs.} \rev{We define a point P to be a LOP in a sub-region $A$ for the sub-expression $S_i$ if it is either the {\color{brown} x.e} of the first \ssg \emph{or} the {\color{brown} x.s} of the last \ssg of  $S_i$.  For instance, in the above example, it is easy to see that a  LOP $P$ in sub-region $A$ is the {\color{brown} x.e} value of [\pup] in the optimal segmentation of [\pup]\colorconcat[\pdown] in A. Since a \concat operation with $k$ operands can have $(k^2)$ sub-sequences, there can be a maximum of $2.k^2$ LOPs in A. The SelectLOPs function (line 9) is used for selecting LOPs. It is a variant of Algorithm~\ref{algo:scoringalgo} that returns both the final score as well as the end points of lines that form the optimal segmentation.}

\stitle{Merging.} \rev{
Next, the algorithm incrementally merges nodes in a  bottom-up fashion to select LOPs over larger sub-regions (lines 6 to 17). More specifically, the Merge function (line 23) merges a sub-sequence $t1: s_i$\colorconcat $s_{i+1}$ ...\colorconcat$s_{i+m-1}$ in the left child with a sub-sequence $t2:s_j$\colorconcat $s_{j+1}$ ...\colorconcat$s_{j+n-1}$ in the right child if (i) $s_{i+1}$...\colorconcat$s_{i+m-1}$\colorconcat$s_j$...\colorconcat
$s_{j+n-1}$ is a subsequence of query $Q$, or (ii) $s_{i+1}$...\colorconcat$s_{i+m-1}$\colorconcat$s_{j+1}$...\colorconcat  $s_{j+n-1}$ is a subsequence of query $Q$ when $s_{i} = s_{j}$ (i.e., we consider the common boundary \ssg only once). While the score of the merged subsequence for case (i) can be easily computed using the average of the scores of left and right subsequence weighted by their number of \ssgs, i.e, $\frac{m\times score(t1) + n\times score(t2)}{m+n}$ , for case (ii) we rescore the common \ssg by estimating the slope of a line from the {\color{brown} x.s} of the last \ssg of $t1$ to {\color{brown} x.e} of the first \ssg of $t2$. If the $s_c$ is the score of common \ssg, $t1^*$ and $t2^*$ are the scores of left and right subsequence without the common \ssg, then the score of the merged sub-sequence is: $\frac{(m-1)\times score(t1^*) + s_c + (n-1)\times score(t2^*)}{m+n-1}$.
When multiple sub-sequences in the children nodes generate the same sub-sequence in the parent node, we select the sub-sequences that result in maximum score after merging, i.e., the one with the best optimal segmentation (line 24--25), thereby pruning out LOPs corresponding to non-selected sub-sequences.
This merging process is repeated at each intermediate node. Finally, at the root node, we select the points that result in the maximum score for the entire sequence of operands.  
 }

\eat{
\vspace{-3pt}
\begin{definition}[\sgt]
	A \sgt over a sub-region $L$ is a balanced binary tree structure, where (i) the leaf nodes correspond to non-intersecting sub-sub-regions of $L$ of size $2$, (ii) the internal nodes cover the concatenation of sub-sub-regions of their children and (iii) the root node covers the entire sub-region $L$. Each node consists of LOPs over its covered sub-sub-region that result in highest scores for sub-sequences of the \concat operation.
\end{definition}
\vspace{-3pt}
}

\begin{figure}
   \vspace{-15pt}
	\centerline {
		\hbox{\resizebox{\columnwidth}{!}{\includegraphics{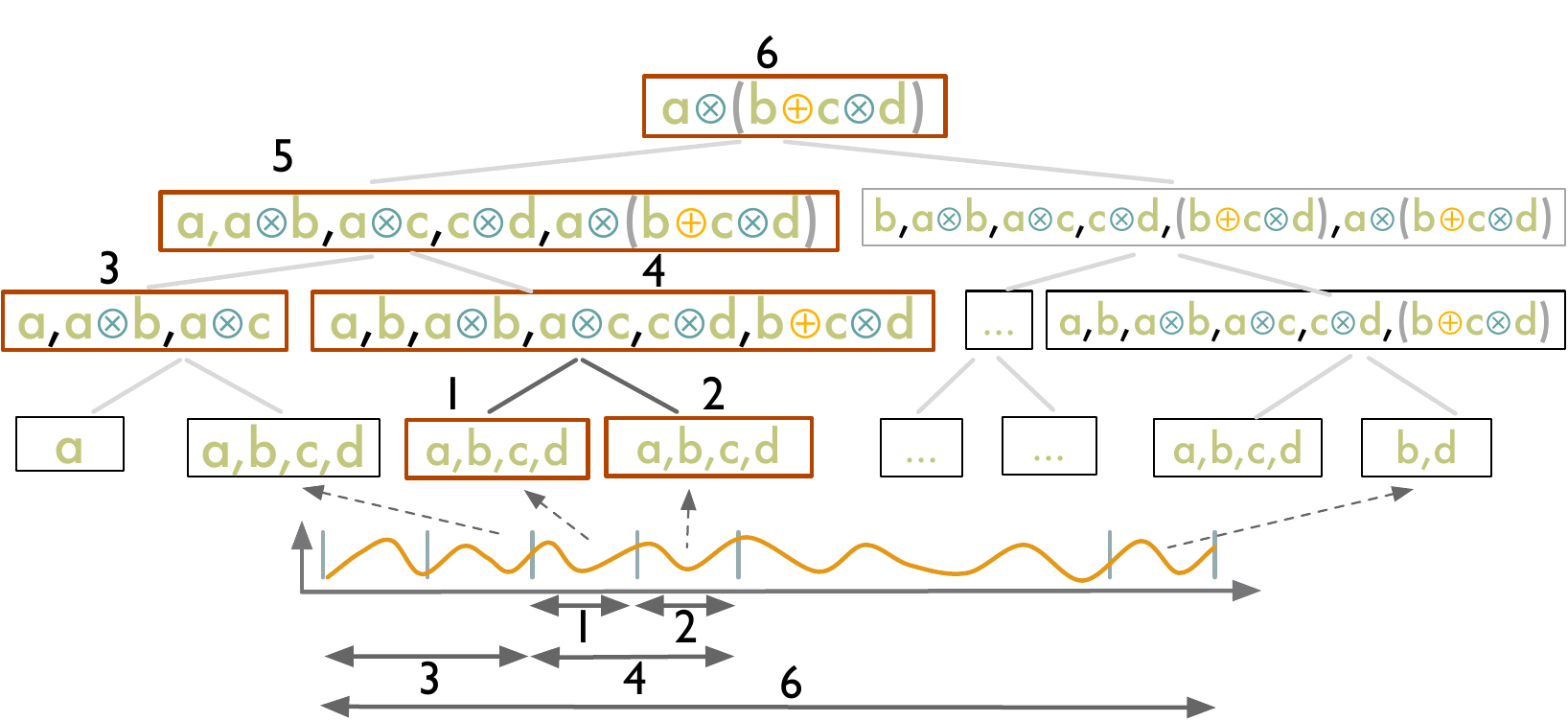}}}}
	\vspace{-8.5pt}
	\caption{\smallcaption{Bottom-up scoring of ShapeQuery}}
	\vspace{-12.5pt}
	\label{fig:segmenttree}
\end{figure}

Figure~\ref{fig:segmenttree} depicts the logical order for scoring \sq \pat{a}\colorconcat(\pat{b}\coloror(\pat{c}\colorconcat\pat{d})) over the sub-sub-regions. Here, \pat{a}, \pat{b}, \pat{c}, and \pat{d} represent a \ssg. 
The \sgt algorithm starts by scoring  individual \ssgs (e.g., \pat{a},\pat{b},\pat{c} and \pat{d}  in \pat{a}\colorconcat(\pat{b}\coloror (\pat{c}\colorconcat\pat{d}))) independently over each of leaf nodes as depicted in Figure~\ref{fig:segmenttree}. Next, it computes the scores of sub-sequences in the intermediate nodes using the merging process described below.
For example, in Figure~\ref{fig:segmenttree}, node 4 depicts the sub-sequences formed by combining sub-sequences from nodes $1$ and $2$, and node 5 depicts the sub-sequences formed by combining sub-sequences from nodes $3$ and $4$.
When multiple sub-sequences in the children nodes generate the same sub-sequence in the parent node, we select the sub-sequences that result in maximum score after concatenation (i.e., the one with the best optimal segmentation), thereby pruning out LOPs corresponding to non-selected sub-sequences.  For example, at node 5,  \pat{a}\colorconcat\pat{b} can be computed from 1) \pat{a} from node $3$ and \pat{b} from node $4$,
2) \pat{a}\colorconcat\pat{b} from node $3$ and \pat{b} from node $4$, and 3)  \pat{a} from node $3$ and  \pat{a}\colorconcat\pat{b} from node $4$. Among the 3 concatenations, we pick the one that gives the maximum score.

\begin{algorithm}
	\small 
	\caption{\rev{Fuzzy Matching Algorithm}}
	\label{algo:segmentreealgo}
	\begin{flushleft}
		\textbf{Input:} $L$: a sub-region of trendline, $Q$: a CONCAT operation, $ScrFunc$: scoring functions from Tables~\ref{tab:scoresgmt} and ~\ref{tab:scoreopt} \\
		\textbf{Output:} score
	\end{flushleft}
	\begin{algorithmic}[1]
		\Procedure{ExecFuzzyQuery}{$L$, $Q$, $ScrFunc$} 
		\State $subRegions$ $\gets$ $ComputeSubRegions(L)$ {\em \color{gray} // leaf nodes} 
		\State $T$ $\gets$ $ComputeSubSequences(Q)$ {\em \color{gray} } 
		\State $nodes$ $\gets$ Queue()
		\State //{ \em scoring of leaf segments}
		\For{each $s$ in $subRegions$}
			\State $lops$ $\gets$ []
			\For{each $t$ in $T$}
				\State $lops$[$t$] $\gets$ SelectLOPs($s$,$t$)
			\EndFor
			\State $node$ $\gets$ [$s$.start, $s$.end, lops[$t$]]
			\State $nodes$.add(node)
		\EndFor	   
		\State // { \em  bottom-up processing}
	    \While{$nodes$.size() > 1}
		\State $s$ $\gets$ $nodes$.Size()
		\State // {\em pairwise merging of nodes at the same level}
		\While {$s$ > 0}
			\State $s1$  $\gets$ $nodes$.deque(), $s2$  $\gets$ $nodes$.deque()
			\State $mlops$ $\gets$ []
			\For{each $t1$,$t2$ in $s1$.$lops$.keys(),$s2$.$lops$.keys()}
				\State $score$, $lops$ $\gets$ Merge($L$, $s1$[$t1$],$s2$[$t2$])
				 \If{$score$ $>$  $mlops$[$t1$\colorconcat$t2$].score}
				\State $mlops$[$t1$\colorconcat$t2$ ] = \{$lops$, $score$\}
				\EndIf
			\EndFor
			\State $node$ $\gets$ [$s1$.start, $s2$.end, $mlops$]
			\State $nodes$.add(node)
			\State $s$ = $s$ $-$ $1$;
		\EndWhile
		\EndWhile
		node $\gets$ $nodes$.deque()
		\State return $node$.$lops$[$Q$].$score$
		
		\EndProcedure
	\end{algorithmic}
\end{algorithm}

\agppapertext{Given the closure assumption, we prove in~\cite{techreport} that the merging process leads to optimal segmentation.}

\vspace{-5pt}
\begin{theorem} Given the closure assumption, the bottom-up algorithm with $k$ \concat operands is optimal with a time complexity of $O(nk^4)$, i.e., linear in the number of points in the trendlines.
\end{theorem}
\vspace{-5pt}

\iftechreportcode
	\smallskip 
\textsc{Proof:} We prove the above theorem via induction.

\emph{Base case.}  For a single node \sgt, there is no difference between the \sgt algorithm and DP, since the \sgt algorithm uses DP to select the LOPs for a single node. 

\emph{Induction step.} Let $L$ and $R$ be two sibling nodes in the \sgt consisting of optimal scores for each possible subsequence of operands in the \concat operations, and let $P$ be their parent node. Let $S^L_{i(k-1)}$
be the score of sub-expression from operand $i$ until $k$ in $L$ for the optimal segmentation of $(i-1)$th to $k$th operands in L and $S^R_{(k+1)j}$ be the score of the sub-expression from operand $k+1$ until $j$ for the optimal segmentation of $k$th to $j+1$th operands in $R$. Let $S^P_{ij}$ be the score of sub-expression  of operand $i$ until $j$ in $P$, formed by concatenation of operands  $i-1$ until $k$ in $L$ and $k$ until $j+1$ in $R$. 
As per the Closure assumption, the optimal segmentation corresponding to $S^P_{ij}$ must include the optimal segmentation $i-1$ until $k$ in $L$ and  $k$ until $j+1$ in $R$.
Since $kth$ operand is common between $L$ and $R$, we need to re-compute its score over the sub-region from \loc{x.e} of $(k-1)$th operand in $L$ and \loc{x.s} of $(k+1)$th operand in $R$ during concatenation. Let $sc^P_k$ be the re-computed score of the $k$th \ssg. Then,  $S_{ij}$ can thus be computed as: 

{	$S_{ij}$
	$=\underset{k}{MAX}\{\frac{(k-i)\times S^L_{i(k-1)})+ sc^P_k + (j-k)\times S^R_{(k+1,j)})}{(j-i+1)}\}$
}

Since, for computing $S_{ij}$, we consider all possible combinations of optimal segmentations in $L$ and $R$ and pick the one that gives the maximum score, it must be optimal. 

\emph{Conclusion.} Thus, by the principle of induction, the \sgt algorithm must also be optimal over the entire \sgt.

\emph{Time Complexity.} For a sub-region of $n$ points, the maximum number of leaf nodes is $n/2$ (since we need at least 2 points per sub-region) and therefore the total number of nodes in the tree is $n$. At each of the leaf node, we estimate the scores of each \ssg independently, taking $O(n\times k)$ operations across all leaf nodes. 
 Each intermediate node involves a merge step, involving concatenation of subsequences from left node with the right node. For $k$ operands in \concat, there can be a maximum of  $k^2$ subsequences per node, requiring a total of  $k^4$ concatenations. Moreover, each concatenation involves the  computation of score $sc^P_k$ of the $k$th \ssg that intersects left and right child. The computation of $sc^P_k$ involves the estimation of the slope of  line from \loc{x.e} of $(k-1)$th \ssg in $L$ to \loc{x.s} of $k+1$th \ssg in $R$, which can be done in constant time from the statistics of $k$th \ssg's sub-region in $L$ and $R$ (see Theorem 3.3). Thus, each merging step involves $O(k^4)$ operations.
Overall, the \sgt algorithm takes $O(n/2\times k^4+nk)$ $\approx$  $O(nk^4)$ time, i.e., linear in the number of points in the sub-region. In practice, $k^4$ is not a problem, since not all combinations of sub-sequences lead to a valid sub-sequence in the \concat operands, therefore the actual number of merges are much fewer. Moreover, $k$ is typically small ($\leq 5$). $\square$.
	
\fi

\vspace{-10pt}
\subsection{Pruning Optimization}
\label{subsec:twostage}
A large number of \sqs are sequential pattern matching queries, consisting of only a single \concat operation on a sequence of simple patterns such as \pat{up}, \pat{down}, \thetx. For such \concat operations, we can bound the  final scores of trendlines and filter low-scoring trendlines without scoring them until the root node of the \sgt. We first describe our key observations.

\stitle{Observation 3.1}
Given a sub-region $L$ comprising of sub-sub-regions: $L_1, L_2, ..., L_n$, the score of a \ssg consisting of patterns \pat{up} or \pat{down} over L, $score_{up/down,L}$  is bounded between the maximum and minimum scores over any of the smaller sub-regions, i.e., $\underset{i}{MIN}(score_{up/down,L_i})$ $\leq$ \\ $score_{up/down,L}$ $\leq$  $\underset{i}{MAX}(score_{up/down,L_i})$. 

This observation holds because the scores of \pat{up} or \pat{down} vary monotonically with the slope of the line, and the slope of the line over the large sub-region is always bounded between the maximum and minimum slopes of the lines over any smaller regions, $\underset{i}{MIN}(slope_{L_i})$ $ \leq$ \\ $slope_{L}$ $\leq$ $\underset{i}{MAX}(slope_{L_i})$

However, when a slope (e.g., {\thetx}) is specified as pattern, the above observation does not hold for $\underset{i}{MIN}(slope_{L_i})$ $\leq$ $x$ $\leq$ $\underset{i}{MAX}(slope_{L_i})$, because the $score_{x,L}$ can be more than $\underset{i}{MAX}(score_{x,L_i})$ when  
$|x - slope_{L}| \le \underset{i}{MIN}(x -slope_{L_i})$. For such cases, we set the upper bound to $1$, the maximum possible score. 
	
\stitle{Observation 3.2}
The score of  an operator is bounded between the minimum and maximum scores of input \ssgs. 

This observation is clear from the scoring functions of operators as defined in Table~\ref{tab:scoreopt}.

Based on the above observations, we can derive the bounds on the final score of a \ssg at the root node using the maximum and minimum scores of the \ssg at a given level $i$ in the \sgt. 
\agptechreport{We summarize the bounds for each of the patterns in Table ~\ref{tab:max_min_scores}.} 

Thus, instead of processing each trendline completely in one go, we process trendlines in rounds. In each round, we process one level of \sgt for all of the trendlines simultaneously, and incrementally refine the upper and lower bounds on their scores. Before moving on to the upper levels, we prune the trendlines that have their upper bound score lower than the current top-$k$ lower bound scores. Overall, the pruning optimization helps avoid processing to completion for a large number of trendlines in the collection, and is particularly effective when the user is looking for trendlines with rare patterns.


\agptechreport{
\begin{table}
	\begin{center}
		\agppapertext{\vspace{-10pt}}
		\caption{\small Bounds on scores for different patterns based on scores at a given level $i$ in the \sgt}
		\vspace{-10.5pt}
		\label{tab:max_min_scores}
		\resizebox{0.95\columnwidth}{!}{%
			\begin{tabular}{ |l|l|l| } 
				\hline
				\textbf{Pattern} & \textbf{Max possible Score} & \textbf{Min possible Score} \\ \hline
				 up & max across all level $i$ nodes & min across all level $i$ nodes\\ \hline
				down & max across all level $i$ nodes & min across all level $i$ nodes\\ \hline
				flat & \pbox{3cm}{max across all level $i$ nodes if all $\theta$ > 0 or  all $\theta$ < 0; otherwise 1}  & min across all level $i$ nodes\\ \hline
				$\theta=x$ & \pbox{3cm}{max across all level $i$ nodes if all $\theta$ > $x$ or  $\theta$ < $x$ ; otherwise 1} & min across all level $i$ nodes \\ \hline
			\end{tabular}%
		}
	\end{center}
	\vspace{-10pt}
\end{table}
}

\begin{table*}[ht]
	\parbox{.37\linewidth}{
		\centering
		\agppapertext{\vspace{-35pt}}
		\caption { \smallcaption{NL Features. d(x) denotes the number of words between current word and x, x+ and x- denote next and previous x}}
		\label {tab:features}
		\resizebox{0.37\textwidth}{!}{%
			\label{tab:features}
			\begin{tabular}{|p{2cm}|p{6.2cm}|} 
				\hline
				Type & Features \\ \hline
				POS Tags &  
				pos-tag, pos-tag-, pos-tag+ \\ \hline
				Words &  word-, word+, word--, word++  \\ \hline
				synonym &  synonym, synonym+, synonym-, d(synonym+), d(synonym-) \\ \hline
				Space and time prepositions & time-preposition+, time-preposition-, space-preposition+, space-preposition-, d(time-preposition+), d(time-preposition-), d(space-preposition+), d(space-preposition-) \\ \hline
				Punctuation &  d(,+), d(,-) ,d(;+), d(;-), d(.+),  d(.-) \\ \hline
				Conjunctions & d(and+), d(or-), d(and then+) \\ \hline
				Miscellaneous & d($x$), d($y$), d(next), ends(ing), ends(ly), length(query) \\ \hline
			\end{tabular}%
		}
	}
	\hfill
	\parbox{.62\linewidth}{
		\centering
			\vspace{-24.5pt}
		\caption{Common Ambiguities and their Resolution}
		\vspace{-10pt}
		\resizebox{0.61\textwidth}{!}{%
			\label{tab:ambg}
			\begin{tabular}{ |p{6.5cm}|p{6.2cm}| }
				\hline
				\textbf{Ambiguity (example queries with predicted entities)}                                                                                                              & \textbf{Rules for Resolution}                                                                                                                              \\ \hline
			
				A1: Conflicting \location and \pattern in a \ssg (e.g., [\underline{decreasing} ($p$) from \underline{$4$} ($x.s$) to \underline{$8$} ($x.e$)])                                                                    & R1: Change the sub-primitive of \location from $x$ to $y$ or $y$ to $x$.  R2: Swap the start and end positions of \location.                            \\ \hline
				A2: Multiple $p$ in the same \ssg
				(e.g., [\underline{increasing} ($p$) from \underline{$2$} ($x.s$) to \underline{$5$} ($x.e$) with \underline{decreasing} ($p$)] \underline{next} (\colorconcat))  & R1: Move one of the $p$s to the adjacent \ssg with missing $p$. R2: split the \ssg into two new \ssgs with an OR operator between them\\ \hline
				A3: Overlapping \ssgs with \colorconcat
				(e.g., \underline{increasing} ($p$) from \underline{$4$} ($x.s$) to \underline{$8$} ($x.e$) \underline{and then} (\colorconcat) \underline{decreasing} ($p$) from \underline{$8$} ($x.s$) to \underline{$0$} ($x.e$) &
				 R1: Change $x$ to $y$, if $y$ values missing. If $y$ values already present, replace {\colorconcat} with {\colorand}  operator.                 \\ \hline
			\end{tabular}%
		}
	}
	\vspace{-10pt}
\end{table*}

\agppapertext{
\stitle{Additional Optimizations.} \ssr minimizes multiple passes over data by reusing the aggregate statistics over smaller sub-regions to estimate the slope of lines over larger sub-regions. In particular, given $\sum x_i$, $\sum y_i$, $\sum x_i.y_i$, $\sum x_i^2$, and $n$ for sub-regions $A$ and $B$) individually, the slope of the line over the combined region $AB$ can be computed as: $\theta_{AB}$=  $\frac{(n_A+n_B)*(\sum x_{Ai}.y_{Ai} + \sum x_{Bi}.y_{Bi}) - (\sum x_{Ai}+\sum x_{Bi})*(\sum y_{Ai}+\sum y_{Bi})}{(n_A+n_B)\times\sum((x_{Ai})^2+(x_{Bi})^2) - \sum(x_{Ai}+x_{Bi})^2}$.

Additionally, when a \sq consists of both fuzzy and non-fuzzy \ssgs, location constraints are pushed down to
the trendline generation component to prune trendlines that do not have any value in the specified $x$ ranges. 
}

\agptechreport{
\vspace{-5pt}
\subsection{Additional Optimizations}
\label{sec:additionaloptm} 

\ssr supports a couple of additional optimizations that result in faster scoring of trendlines.

\stitle{Generating lines via Summary Statistics}. For scoring a sub-region, \ssr fits a line to approximate it. This is costly for fuzzy \sqs where \ssr needs to score sub-regions of varying sizes, fitting one line for every sub-region. We note that a summary of five statistics namely, $\sum x_i$,  $y_i$, $\sum x_i.y_i$, $\sum x_i^2$, and $n$ for a sub-region, is sufficient to  compute the slope of the line over the sub-region as follows:
$\theta$=$\frac{(n\times\sum x_i.y_i - \sum x_i\sum y_i)}{ (n\times\sum x_i^2 - (\sum x_i)^2)}$, $\delta$=$\sum y_i-\theta\times\sum x_i$.\\
Moreover, it is easy to see that the individual summaries over two sub-regions ($A$ and $B$) are sufficient to compute the slope of the line over the combined region $AB$, without making additional passes over the data.

$\theta_{AB}$=  $\frac{(n_A+n_B)*(\sum x_{Ai}.y_{Ai} + \sum x_{Bi}.y_{Bi}) - (\sum x_{Ai}+\sum x_{Bi})*(\sum y_{Ai}+\sum y_{Bi})}{(n_A+n_B)\times\sum((x_{Ai})^2+(x_{Bi})^2) - \sum(x_{Ai}+x_{Bi})^2}$ \\

Thus, the summary statistics help reduce data movement as well as the amount of data  processed during segmentation. 
\agptechreport{
We summarize our finding using the following theorem.
\begin{theorem}[Additivity] Given two adjacent segments A and B, a line segment over the combined segment AB can be estimated using linear regression on the summarized statistics over the individual segments A and B.
\end{theorem}
}

\stitle{Push-Down Optimizations}. 
\ssr applies a number of push-down optimizations when a \sq involves location constraints. Consider a \sq: [{\color{orange} p= up,x.s=50, x.e=100}][\pdown][\pup] that searches for shapes which are  $increasing$  from $50$ to $100$ followed by a decreasing, and then an $increasing$ pattern. \ssr employs three push-down optimizations for such queries: \textbf{(1)} \location primitives in \sq are pushed down to the trendline generation component to prune trendlines that do not have any value in the specified $x$ ranges (e.g., $50$ to $100$ in the above query), \textbf{(2)} When a \sq contains a \ssg with an {\pup} or {\pdown}  pattern along with both start and end locations (e.g., [\pup, {\color{orange} x.s=50,x.e=100}] in the above query), \ssr prioritizes the  segmentation of \ssgs over such location primitives first, since the trendlines with negative scores over such sub-regions tend to have substantially lower scores. This helps the pruning of low scoring trendlines much earlier in the \sgt, and \textbf{(3)} Finally, \ssr avoids computing \emph{summary statistics} over $x$ ranges that are not used in the \sq (e.g., $0$ to $50$ in the above query), since the values over such ranges are ignored for segmentation and scoring.  Overall, as we will see in Section~\ref{sec:exp}, these push-down optimizations significantly help in improving the overall response time of the \ssr.
 
}


\section{Natural Language Translation}
\label{sec:translation}

So far, we haven't described how natural language
queries are parsed into \sqs. 
\agppapertext{
We provide a brief overview of the three key steps involved in
parsing, and refer readers to our extended report~\cite{techreport}
for additional details. \rev{We use the following natural language query collected from MTurk for illustration: show me the trendlines that are increasing from $2$ to $5$  and then decreasing}

\stitle{Step 1. Primitives and Operators Recognition.} 
Given a natural language query, 
the first step is to map words to their corresponding 
shape primitives and operators. 
\rev{For example, the above query is tagged as 
``show (noise) me (noise) the (noise) trendlines (noise) that (noise) are (noise) \underline{increasing} ({\color{olive} p}) \underline{from} (noise) $2$ ({\color{orange} x.s}) \underline{to} (noise) \underline{$5$} ({\color{orange} x.e}) \underline{and then} (\colorconcat)  \underline{decreasing} ({\color{olive} p})''.}
In order to do so, we learn a linear-chain conditional-random field model (CRF)~\cite{lafferty2001conditional} 
and train it on the same $250$ natural language queries 
we collected via Mechanical Turk (described in ~\cite{techreport}) for understanding query characteristics, 
as alluded to in Section~\ref{sec:intro}. For each word, we use its part-of-speech (POS) tags 
along with word-level features 
as outlined in Table~\ref{tab:features}. 

\stitle{Step 2. Identifying Pattern Value.} 
\rev{For each of the words predicted of type {\color{olive} p}, e.g., increasing and decreasing in the above query, we additionally map them to the corresponding semantic pattern supported in \ssr, e.g., ``increasing'' is mapped to \pup}. For this mapping, \ssr computes the similarity between the specified word and synonyms of the supported patterns, first using edit distance and then using wordnet~\cite{roetter2005integration}. The semantic pattern with the highest similarity between any of its synonyms and the specified word is selected. 


\stitle{Step 3. ShapeQuery Generation and Ambiguity Resolution.} Next, we group primitives and operators into a \sq. \ssr first groups all the primitives between two operators into a single \ssg. For instance,  for the above query, the primitives are grouped as follows: \rev{[\underline{increasing} (\pup),  $2$ ({\color{orange} x.s}), \underline{$5$} ({\color{orange} x.e}) ]  \underline{and then} (\colorconcat) [ \underline{decreasing} (\pdown)].}  In some cases, this may lead to  incorrect grouping of primitives, e.g., two patterns in the same \ssg. Moreover, there could be semantic ambiguity because of incorrect entity tagging, e.g., \underline{decreasing} (\pup) \underline{from} \underline{$5$} ({\color{orange} y.s}) to \underline{$10$} ({\color{orange} y.e})  where {\color{orange} x.s} and {\color{orange} x.e} values are wrongly tagged as {\color{orange} y.s} and {\color{orange} y.e} respectively. \ssr uses rule-based transformations that try to reorder and change the types of entities to get a correct and meaningful \sq. \rev{In Table~\ref{tab:ambg}, we list three common ambiguities (A1, A2, A3) with real examples from mturk corpus and a sequence of rules (e.g., R1, R2) that are applied in order to resolve these.}

The parsed \sq is sent to the front-end, and displayed as part of the correction panel (Box 4 in Figure \ref{fig:interface}) for users to edit or further refine the parsed representation if needed. The validated query is then executed to generate the matching trendlines.
}

\agptechreport{
	We now provide a brief overview of the key steps involved in
	parsing. \rev{We use the following natural language query collected from MTurk for illustration: ``show me the trendlines that are increasing from $2$ to $5$  and then decreasing''}
	
	\stitle{Step 1. Primitives and Operators Recognition.} 
	Given a natural language query, 
	the first step is to map words to their corresponding 
	shape primitives and operators. 
	We follow a two-step process. First, using the Part-of-Speech (POS) tags and word-level features, we classify each word in the query as either noise or non-noise. For example, words $\in$ \{determiner, preposition\, stop-words\} are more likely to be noise, while words $\in$ \{noun, adjective, adverb, number, transition words, conjunction\} may refer to a primitive or operators. Next, given a sequence of non-noise words, we use a linear-chain conditional-random field model (CRF)~\cite{lafferty2001conditional} (a probabilistic graphical model used for modeling sequential data, e.g., POS tagging) to predict their corresponding primitives and operator. 
	\rev{For example, the above query is tagged as 
		``show (noise) me (noise) the (noise) trendlines (noise) that (noise) are (noise) \underline{increasing} ({\color{olive} p}) \underline{from} (noise) $2$ ({\color{orange} x.s}) \underline{to} (noise) \underline{$5$} ({\color{orange} x.e}) \underline{and then} (\colorconcat)  \underline{decreasing} ({\color{olive} p})''.}.
	
	We train the CRF model~\cite{lafferty2001conditional}  on the same $250$ natural language queries 
	that we used for characterizing trendline patterns (Section~\ref{sec:intro}). We provide more details on how we collected the queries in Appendix~\ref{sec:mturk}.
	We extract a set of features (listed in Table~\ref{tab:features}) for each non-noise word in the sequence. 
	In addition, \ssr stores ``synonyms'' for each primitve and operator (e.g., ``increasing" for up, ``next" for \concat),  and if a non-noise words closely matches with them (e.g., with edit distance $<= 2$), we add the matched primitive or operator as a feature called \emph{predicted-entity}. This idea is inspired from the concept of ``boostrapping'' in weakly-supervised learning approaches~\cite{kozareva2011class,siddiqui2016facetgist}, and helps improve the overall accuracy. We implemented the model using the Python CRF-Suite library ~\cite{pythoncrfsuite} with parameter settings: \emph{L1 penalty:1.0, L2 penalty:0.001, max iterations: 50,  feature.possible-transitions: True}.
	On $5$-fold cross-validation over the crowd-sourced queries, the model had an $F1$ score of $81\%$ ($precis on=73\%$, $recall=90\%$).
	
	\stitle{Step 2. Identifying Pattern Value.} 
	\rev{For each of the words predicted of type {\color{olive} p}, e.g., increasing and decreasing in the above query, we additionally map them to the corresponding semantic pattern supported in \ssr, e.g., ``increasing'' is mapped to \pup}. For this mapping, \ssr computes the similarity between the specified word and synonyms of the supported patterns, first using edit distance and then using wordnet~\cite{roetter2005integration}. The semantic pattern with the highest similarity between any of its synonyms and the specified word is selected. 
	
	\stitle{Step 3. ShapeQuery Generation and Ambiguity Resolution.} Next, we group primitives and operators into a \sq. \ssr first groups all the primitives between two operators into a single \ssg. 
	For instance,  for the above query, the primitives are grouped as follows: \rev{[\underline{increasing} (\pup),  $2$ ({\color{orange} x.s}), \underline{$5$} ({\color{orange} x.e}) ]  \underline{and then} (\colorconcat) [ \underline{decreasing} (\pdown)].}  
	In some cases, this may lead to  incorrect grouping of primitives, e.g., two patterns in the same \ssg. Moreover, there could be semantic ambiguity because of incorrect entity tagging, e.g., \underline{decreasing} (\pup) \underline{from} \underline{$5$} ({\color{orange} y.s}) to \underline{$10$} ({\color{orange} y.e})  where {\color{orange} x.s} and {\color{orange} x.e} values are wrongly tagged as {\color{orange} y.s} and {\color{orange} y.e} respectively. \ssr uses rule-based transformations that try to reorder and change the types of entities to get a correct and meaningful \sq.  In Table~\ref{tab:ambg}, we list three common ambiguities (A1, A2, A3) and a sequence of rules (e.g., R1, R2) that are applied in order to resolve these.
	
	The parsed \sq is sent to the front-end, and displayed as part of the correction panel (Box 4 in Figure \ref{fig:interface}) for users to edit or further refine the parsed representation if needed. The validated query is then executed to generate the matching trendlines.
}


\agppapertext{
\begin{table}
	\centering
	\caption{\smallcaption{Real-world Datasets and Query Characteristics}}
	\label{tab:dataset}
	\vspace{-13.5pt}
	\resizebox{\columnwidth}{!}{%
		\begin{tabular}{l|p{0.5cm}|p{0.5cm}|p{8cm}}

			Name   &  |V|   &  $|V_i|$ & Queries  \\ \hline \hline
			Weather     & 144 & 366 & ({\color{olive} ${\theta}{=} 45^{\circ}$}\colorconcat{\color{olive} d}\colorconcat{\color{olive} u}\colorconcat{\color{olive} d}),
			(({\color{olive} u}\coloror{\color{olive} d})\colorconcat{\color{olive} f}\colorconcat{\color{olive} u}\colorconcat{\color{olive} d}),
			({\color{olive} f}\colorconcat{\color{olive} u}\colorconcat{\color{olive} d}\colorconcat{\color{olive} f} ) 	\\ \hline
			Worms       & 258 & 900 &  
				({\color{olive} d}\colorconcat({\color{olive} ${\theta}{=}45^{\circ}$}\coloror {\color{olive} ${\theta}{=}{-}20^{\circ}$})\colorconcat{\color{olive} f}),
				({\color{olive} d}\colorconcat{\color{olive} ${\theta}{=}45^{\circ}$}\colorconcat{\color{olive} d}),
				({\color{olive} u}\colorconcat{\color{olive} d}\colorconcat{\color{olive} u})
			 	 \\ \hline
			50 Words    & 905 & 270 & 
			 ({\color{olive} d}\colorconcat({\color{olive} u}\coloror({\color{olive} f}\colorconcat{\color{olive} d}))), ({\color{olive} f}\colorconcat{\color{olive} u}\colorconcat{\color{olive} d}\colorconcat{\color{olive} f}), (({\color{olive} {\color{olive} u}}\coloror {\color{olive} d})\colorconcat({\color{olive} u}\coloror{\color{olive} d})\colorconcat{\color{olive} f})
			\\ \hline
			Real Estate & 1777 & 138 &  ({\color{olive} f}\colorconcat{\color{olive} d}\colorconcat{\color{olive} u}\colorconcat{\color{olive} f}),({\color{olive} u}\colorconcat{\color{olive} d}\colorconcat{\color{olive} u}\colorconcat{\color{olive} f}), ({\color{olive} u}\colorconcat{\color{olive} f}\colorconcat(({\color{olive} ${\theta}{=}45^{\circ}$}\colorconcat{\color{olive} ${\theta}{=}60^{\circ}$}) \coloror({\color{olive} u}\colorconcat{\color{olive} d})))
		 \\ \hline
			Haptics     & 463 & 1092 &  ({\color{olive} u}\colorconcat{\color{olive} d}\colorconcat {\color{olive} f}\colorconcat{\color{olive} u}), ({\color{olive} d}\colorconcat{\color{olive} u}\colorconcat{\color{olive} d}\colorconcat {\color{olive} f})
	    	 \\
		\end{tabular}%
	}
	\vspace{-16.5pt}
\end{table}
}

\agptechreport{
	\begin{figure*}[ht]
		\begin{minipage}{.72\linewidth}
			\centering
			\papertext{\vspace{-15pt}}
			\resizebox{\columnwidth}{!}{%
				\label{tab:runtimeaccuracy}
				\begin{tabular}{p{0.08cm} p{0.95cm} p{0.31cm} p{0.32cm} p{2.3cm} | p{1.35cm} p{0.35cm} p{0.55cm} p{1cm} p{1.35cm} p{0.9cm}| p{1cm} p{1cm}|}
					& & & & &  \multicolumn{6}{c|}{Runtime (sec)} & \multicolumn{2}{c|}{Accuracy (\%)}\\ 		
					& Dataset &  |V|   &  $|V_i|$ & Query  & Exhuastive & DP & DTW & Segment Tree & Segment Tree+Prune & Greedy & Segment Tree & Greedy\\ \hline	
					1 & Weather & 144 & 366 &  ({\color{olive} ${\theta}{=} 45^{\circ}$}\colorconcat{\color{olive} d}\colorconcat{\color{olive} u}\colorconcat{\color{olive} d})   & 290 & 52 & 11 & 5 & 2.8 & 0.9 & 85 & 25							\\ 
					2 & Weather & 144 & 366 &  (({\color{olive} u}\coloror{\color{olive} d})\colorconcat{\color{olive} f}\colorconcat{\color{olive} u}\colorconcat{\color{olive} d})  & 211 & 55 & 9 & 4 & 3.2 & 1.1 & 90 & 30		 \\ 
					3 & Weather &  144 & 366 & ({\color{olive} f}\colorconcat{\color{olive} u}\colorconcat{\color{olive} d}\colorconcat{\color{olive} f} )   & 244 & 47 & 9 & 5 & 3.3 & 1.4 & 100 & 25		 \\ 
					4 & Worms & 258 & 900 &  ({\color{olive} d}\colorconcat({\color{olive} ${\theta}{=}45^{\circ}$} \coloror {\color{olive} ${\theta} {=}{-}20^{\circ}$} )\colorconcat{\color{olive} f})  & 4737 & 76 & 53 & 10 & 7 & 2.2 & 90 & 35		 \\
					5 & Worms & 258 & 900 &  ({\color{olive} d}\colorconcat{\color{olive} ${\theta}{=}45^{\circ}$}\colorconcat{\color{olive} d})  & 4320 & 63 & 44 & 12 & 9 & 3.4 & 90 & 35		 \\ 
					6 & Worms & 258 & 900 &  ({\color{olive} u}\colorconcat{\color{olive} d}\colorconcat{\color{olive} u})  & 3953 & 68 & 42 & 9 & 6 & 2.5 & 90 & 20		 \\ 
					7 & 50Words & 905 & 270 & ({\color{olive} d}\colorconcat({\color{olive} u}\coloror({\color{olive} f}\colorconcat{\color{olive} d})))  & 1046 & 105 & 28 & 7 & 5 & 1.1 & 90 & 25		 \\ 
					8 & 50Words & 905 & 270 &  ({\color{olive} d}\colorconcat{\color{olive} ${\theta}{=}45^{\circ}$}\colorconcat{\color{olive} d})  & 954 & 122 & 32 & 7 & 5 & 1.9 & 100 & 40		 \\
					9 & 50Words & 905 & 270 &  (({\color{olive} {\color{olive} u}}\coloror {\color{olive} d})\colorconcat({\color{olive} u}\coloror{\color{olive} d})\colorconcat{\color{olive} f}) & 979 & 131 & 29 & 9 & 7 & 1.2 & 85 & 40		 \\ 
					10 & Housing & 1777 & 138 &   ({\color{olive} f}\colorconcat{\color{olive} d}\colorconcat{\color{olive} u}\colorconcat{\color{olive} f}) & 165 & 58 & 40 & 14 & 12 & 1.5 & 80 & 15		 \\ 
					11 & Housing & 1777 & 138 &  ({\color{olive} u}\colorconcat{\color{olive} d}\colorconcat{\color{olive} u}\colorconcat{\color{olive} f})  & 152 & 63 & 41 & 17 & 13 & 1.9 & 85 & 20		 \\ 
					12 & Housing & 1777 & 138 &  ({\color{olive} u}\colorconcat{\color{olive} f}\colorconcat(({\color{olive} ${\theta}{=}45^{\circ}$}\colorconcat  { \color{olive} ${\theta}{=}60^{\circ}$})\coloror({\color{olive} u}\colorconcat{\color{olive} d})))  & 157 & 52 & 35 & 18 & 14 & 1.2 & 75 & 15		 \\ 	
					13 & Haptics & 463 & 1092 & ({\color{olive} u}\colorconcat{\color{olive} d}\colorconcat {\color{olive} f}\colorconcat{\color{olive} u})  & 6869 & 890 & 62 & 16 & 12 & 3.1 & 90 & 40		 \\ 
					14 & Haptics & 463 & 1092 & ({\color{olive} d}\colorconcat{\color{olive} u}\colorconcat{\color{olive} d}\colorconcat {\color{olive} f}) & 7189 & 924 & 58 & 20 & 15 & 2.6 & 95 & 25		 \\ 
				\end{tabular}%
			}
			\caption { \smallcaption{\revtr{Runtime and accuracy results}}}
			\label{fig:perftab}
		\end{minipage}
		\hfill
		\begin{minipage}{.23\linewidth}
			\centering
			\resizebox{\linewidth}{!}{\includegraphics{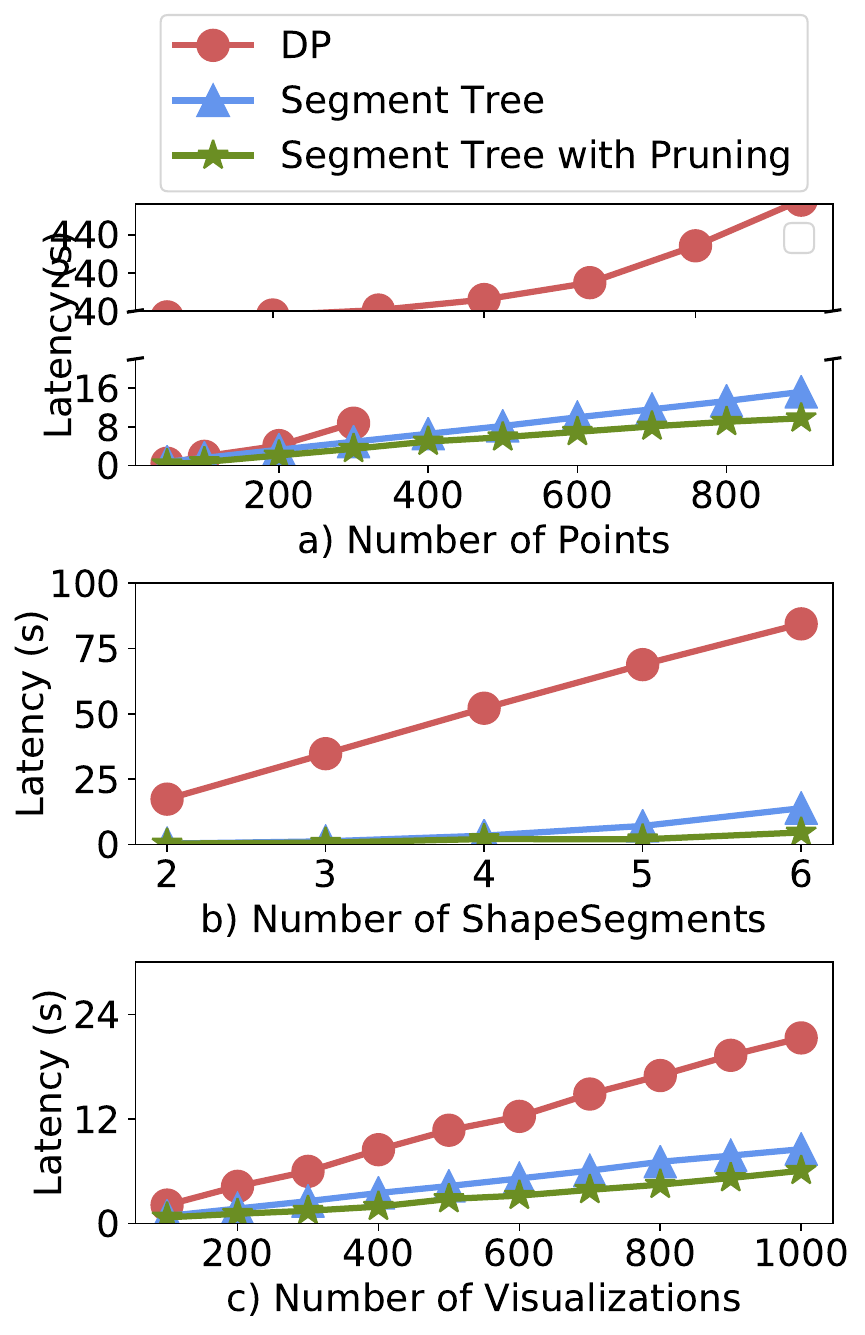}}
			\vspace{-20pt}
			\caption{\small \revtr{Impact on varying characteristics of ShapeQueries}}
			\label{fig:varying-char-merged}
		\end{minipage}
		\vspace{-10pt}
	\end{figure*}
}

\section{Performance Evaluation}
\label{sec:exp}

\eat{We first evaluate the effectiveness of \ssr scoring functions compared to shape similarity measures used in Zenvisage and Qetch using the ground-truth results on the user-study tasks.
} 
In this section, we evaluate the runtime 
and accuracy of \ssr pattern matching algorithms.
 We first compare the runtime of the exhaustive 
 pattern matching algorithm (Section~\ref{sec:scoring}) 
 with four algorithms proposed in Section~\ref{sec:fuzzychallenge}: 
 (i) the dynamic programming-based (DP) algorithm, 
 (ii) the Greedy algorithm, 
 (iii) the \sgt algorithm, and (iv) the \sgt algorithm with pruning.
 We also compare with Dynamic Time Warping 
 (DTW)~\cite{dtw}, another dynamic-programming algorithm that is typically used for matching shapes in trendlines in systems like Zenvisage~\cite{zenvisagevldb}, to show the efficiency of \ssr relative to existing systems. 
 \rev{Next, we compare the accuracy 
 of \sgt and Greedy with respect to the results of DP. Note that \sgt and Greedy are approximate while DP is an optimal algorithm and gives the same results as that of the exhaustive algorithm.}
 Finally, we vary the characteristics 
 of \sqs to assess 
 the impact of different factors on performance. 
 \agppapertext{In our technical report~\cite{techreport}, 
 we also assess the impact of push-down 
 and two-stage collective pruning optimizations from Section~\ref{subsec:twostage}.}

 
\vspace{2pt}
\noindent\textbf{Datasets and Setup.} 
\agppapertext{Table~\ref{tab:dataset}}
\revtr{Figure~\ref{fig:perftab}} depicts the five real-world datasets drawn from the UCI repository~\cite{ucirepo}, and the list of queries we used for our experiments. Each dataset 
consists of trendlines with a mix of shapes,  and the datasets differ from each other in terms of number of trendlines ($|V|$) as well as their length ($|V_i|$). The queries were selected to 
have at least $20$ trendlines with scores  $>0$ 
to ensure that the issued \sqs were relevant to the dataset. 
All experiments were conducted on a 64-bit Linux server with 16 2.40GHz Intel Xeon E5-2630 v3 8-core processors and 128GB of 1600 MHz DDR3 main memory. 
Datasets were stored in memory, and 
we ran six trials for each query on each dataset.

\agppapertext{
\begin{figure}
	\hspace{-0.3cm}
	\centerline {
		\hbox{\resizebox{\columnwidth}{!}{\includegraphics{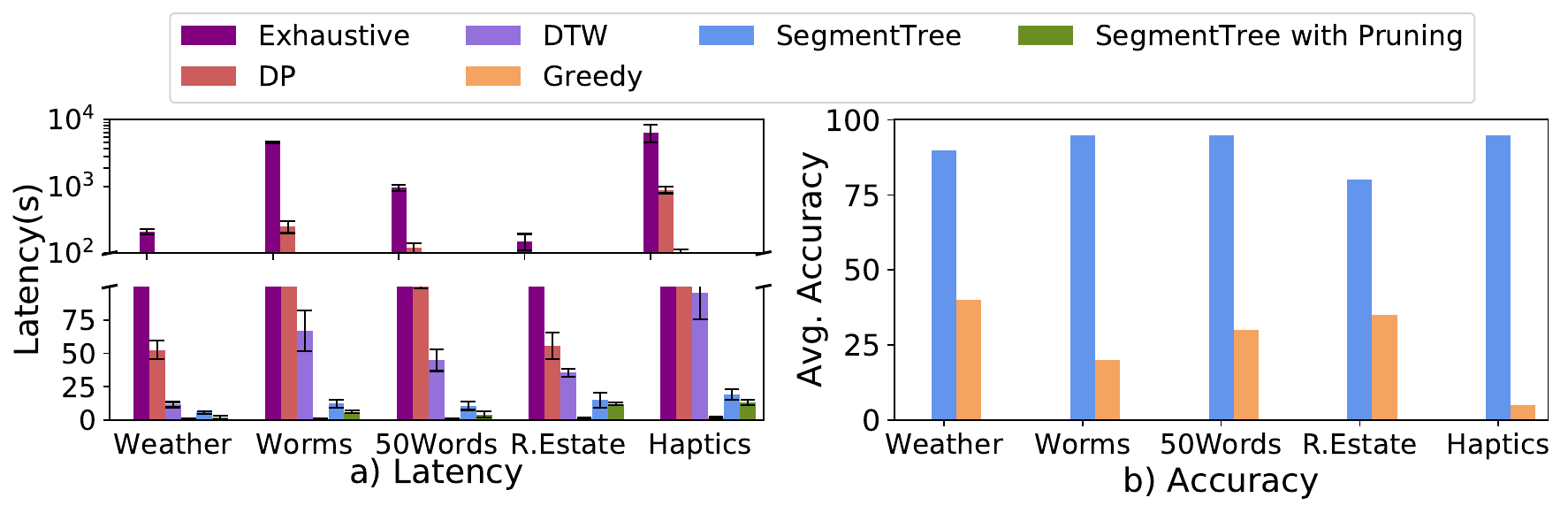}}}}
	\vspace{-10.5pt}
	\caption{\smallcaption{\agppapertext{Runtime and accuracy results.}}}
	\label{fig:perf}
   \vspace{-16.5pt}
\end{figure}
}

\agptechreport{
\begin{figure}
	\centerline {
		\hbox{\resizebox{0.90\columnwidth}{!}{\includegraphics{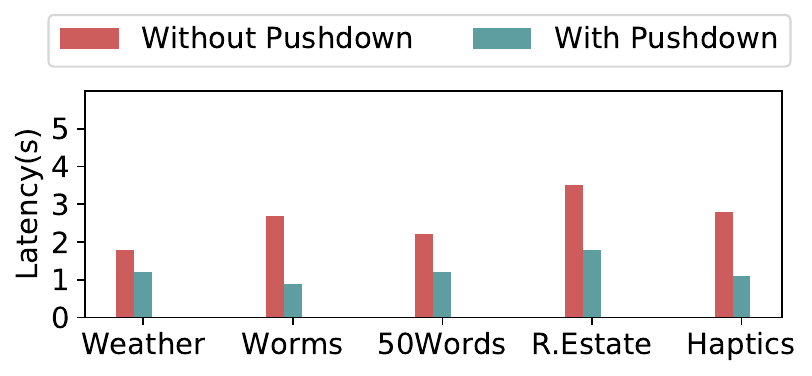}}}}
	\vspace{-10.5pt}
	\caption{\smallcaption{\agptechreport{Average running time before and after push-down optimizations on non-fuzzy queries.}}}
	\label{fig:runtimepushdownopt}
	\vspace{-15.5pt}
\end{figure}
}

\agptechreport{
\begin{figure*}
	\centerline {
		\hbox{\resizebox{\textwidth}{!}{\includegraphics{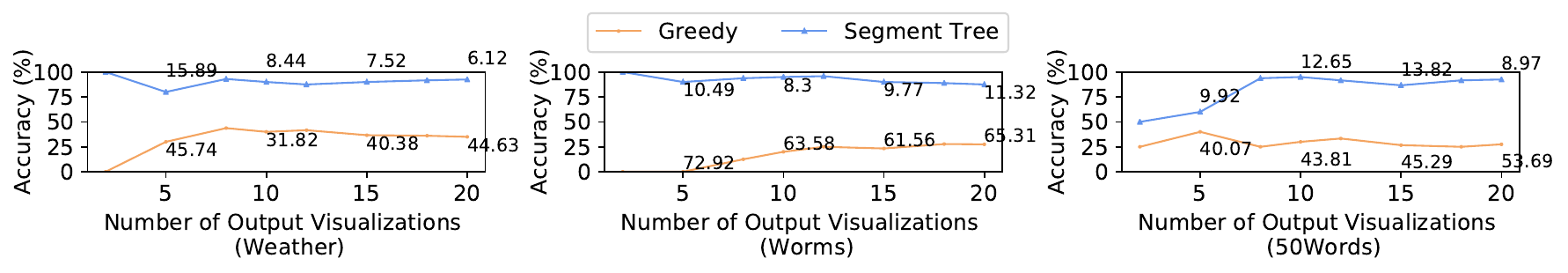}}}}
	\vspace{-10.5pt}
	\caption{\smallcaption{\agptechreport{Accuracy with respect to DP over $3$ real datasets with varying number of output trendlines. Annotations denote the average deviation (in \%) of the score of $kth$ trendline chosen by algorithms with respect to the $kth$  optimal trendline.}}}
	\label{fig:accuracy-indiv}
	\vspace{-2.5pt}
\end{figure*}
}

\agppapertext{
\begin{figure*}
	\vspace{-1.5pt}
	\centerline {
		\hbox{\resizebox{\textwidth}{!}{\includegraphics{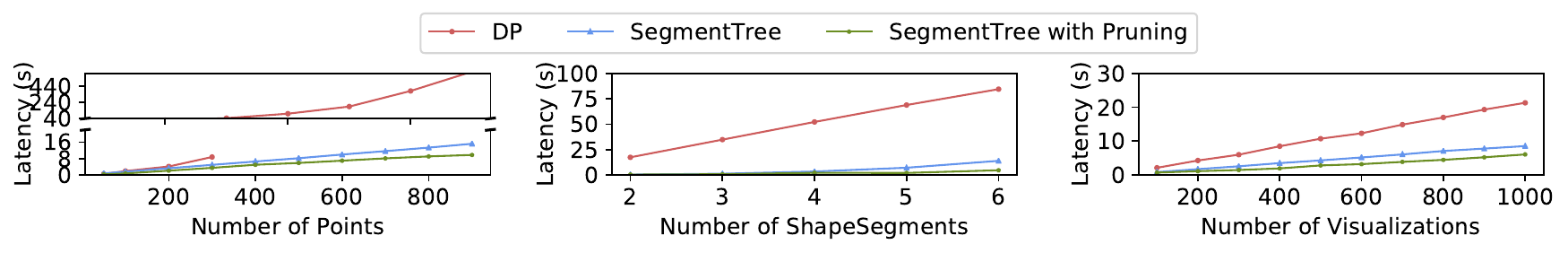}}}}
	\vspace{-10.5pt}
	\caption{\smallcaption{\agppapertext{Runtimes on varying characteristics of ShapeQueries}}}
	\label{fig:varying-char-merged}
	\vspace{-8.5pt}
\end{figure*}
}

\vspace{-5pt}
\subsection{Overall Run-time and Accuracy}
\label{subsec:overall}

 \vspace{-3pt}
\stitle{Runtime Comparison.} 
\agppapertext{Figure~\ref{fig:perf}a depicts the average runtime 
for each of the datasets, 
with the confidence interval indicating 
the maximum and minimum runtime of the algorithm 
across all queries. We show the results for individual queries in our technical report~\cite{techreport}}
\revtr{Figure~\ref{fig:perftab} (Runtime) depicts the runtime for each of the queries across all datasets.}
We see the time taken by the exhaustive algorithm 
is prohibitively large, rendering it no longer interactive. 
DP provides an order of magnitude speed-up over the exhaustive approach; 
however even DP can take $100$s of seconds over trendlines with only a few hundred points. 
Both Greedy and \sgt
provide a $2 \times$ to $40 \times$ improvement in runtime 
compared to DP, taking only a few seconds in the worst case. 
These algorithms explore a much fewer number of 
segmentations compared to the DP approach.  
We also see that these algorithms are about $10 \times$ faster than the DTW algorithm, 
whose runtime, like DP, varies quadratically with the number of points in the trendline. 
Finally, \sgt with Pruning further provides a speed-up of 10-30\% 
by pruning low utility trendlines.
Since the improvement in performance of \sgt and Greedy 
comes at the cost of accuracy, we next compare the accuracies.

\stitle{Acccuracy Comparison.}
\agppapertext{Figure~\ref{fig:perf}b depicts the 
accuracy of \sgt and Greedy relative to DP.}
\revtr{Figure~\ref{fig:perftab} (Accuracy) depicts the 
accuracy of \sgt and Greedy relative to DP.}
We do not compare the accuracy of DTW 
with \ssr algorithms since their scoring functions differ; 
instead we  perform an user study  in the next section to compare 
the effectiveness of \ssr scoring functions 
with DTW and other similar metrics. 
\rev{We define accuracy here to be the number of trendlines 
picked by the algorithm that are also present in the top $20$ trendlines 
selected by DP.}
We see that Greedy  has a low accuracy ($< 30\%$),
since it gets stuck at local optima. 
The accuracy of \sgt
is closer to that of DP and is never off by more than $2$ trendlines when we look at top 10 visualizations.
Unlike Greedy, \sgt compares the local patterns 
in the trendlines 
and those specified in the \sq to select the segmentations that could result in high score. 

\agptechreport{
Figure~\ref{fig:accuracy-indiv} depicts the accuracy results over top-$k$ visualizations (with $k$ varying from $2$ to $20$) for 3 of the the datasets. Annotations in each of the figures depict the average deviation in \% of the score of $kth$ visualization that an algorithm selects with respect to the score of the  $kth$ optimal visualization, indicating how off the shapes of selected visualizations are from optimal ones.  We note that the accuracy of \sgt improves as the number of output visualizations increases, and is never off by more than $2$ visualizations or have more than $> 12\%$ deviation in scores when we look at top $20$ visualizations.}

Overall, the runtime and accuracy results demonstrate that the \textbf{\sgt achieves comparable accuracy to that of DP in much less time}.
 
\agptechreport{
Next, we explore the impact of push-down optimizations, discussed in Section~\ref{sec:additionaloptm}, on the overall performance of queries.
 
\stitle{Impact of Push-Down Optimizations.} We issue non-fuzzy queries, one query for each of the datasets, as depicted in Table~\ref{tab:dataset}.  Figure~\ref{fig:runtimepushdownopt} depicts the runtimes for \ssr (note that all \ssr algorithms behave similarly for non-fuzzy queries) with and without push-down optimizations. We observe that non-fuzzy queries execute very quickly ($<4s$ for over $1000$ trendlines with more than $1000$ points each), but pushdown optimizations help in further reduction of runtime in proportion to the selectivity of the \location primitives in the query. For example, for \sq [\pup, {\color{orange} x.s=60,x.e=80}] on the Haptics dataset, pushdown optimizations help reduce the runtime from $3s$ to  $<1.2s$.
}

\agptechreport{
	\begin{table}
		\centering
		\caption{\smallcaption{Non-Fuzzy Queries}}
		\label{tab:dataset}
		\vspace{-12.5pt}
		\resizebox{\columnwidth}{!}{%
			\begin{tabular}{|l|p{8cm}|}
				\hline
				Name   &  Non-Fuzzy Queries \\ \hline
				Weather    & 
				[{\color{olive}p=down},{\color{brown}x.s=1,x.e=4}]\colorconcat[{\color{olive}p=up},{\color{brown}x.s=4,x.e=10}]\colorconcat[{\color{olive} p=down}, {\color{brown} x.s=10,x.e=12]}						\\ \hline
				Worms      
				&	[{\color{olive} p=down},{\color{brown} x.s=50,x.e=100}]	 \\ \hline
				50 Words   
				&
				[{\color{olive} p=down},{\color{brown} x.s=200,x.e=400}]\colorconcat		[{\color{olive} p=up},{\color{brown} x.s=800,x.e=850}]
				\\ \hline
				Real Estate 
				&	
				[{\color{olive} p=down},{\color{brown} x.s=1,x.e=20}]\colorconcat[{\color{olive} p=up},{\color{brown} x.s=20,x.e=60}] \colorconcat[{\color{olive} p=down},{\color{brown} x.s=60,x.e=138}] 		\\ \hline
				Haptics    
				& [{\color{olive} p=up},{\color{brown} x.s=60,x.e=80}]	 \\ \hline
			\end{tabular}%
		}
		\vspace{-1.5pt}
	\end{table}
}

\vspace{-5pt}
\subsection{Varying \sq Characteristics}
\label{sec:exp-appendix}

We evaluated the efficacy of our \sgt-based optimizations with respect to three different characteristics of \sqs, as discussed below.

\stitle{Impact of number of data points}.
Figure~\ref{fig:varying-char-merged} shows the performance of algorithms as we increase the number of data points in trendlines for a fuzzy \sq (u\colorconcat d\colorconcat u \colorconcat d). With the increase in data points, the overall runtimes increases for all algorithms because of the increase in the number of segmentations. Nevertheless, \sgt shows better performance than DP after 100 data points since the \sgt approach is less sensitive (linear time) to the number of data points than that of DP (quadratic).

\stitle{Impact of number of patterns.}  Figure~\ref{fig:varying-char-merged} depicts the performance of fuzzy \sqs with varying the number of \ssgs (alternating $up$ and $down$ patterns) and issued over the weather dataset. As the number of \ssgs in the \sq grows, the overall runtimes of the algorithms also increases, with the runtimes for \sgt and \sgt with pruning growing much faster ($k^4$) than DP ($k$). However, the overall time for DP is still larger because the number of data points ($366$ in the weather dataset) plays a more dominant ($n^2$) role.

\stitle{Impact of number of trendlines.} 
We increased the number of trendlines from $100$ to $1000$ in the real-estate dataset with a step size of 100 and issued a fuzzy \sq ($u$\colorconcat$d$ \colorconcat $u$\colorconcat$d$); the results are depicted in Figure~\ref{fig:varying-char-merged}. While the overall runtime for all approaches grows linearly with the number of trendlines, the gap between \sgt and \sgt with pruning grows wider. This is because more trendlines get pruned  as the size of the collection grows larger.

\section{User Study}
\label{sec:userstudy}

We conducted a user study 
to perform a 
qualitative and quantitative comparison of 
\ssr with two baseline tools: our prior work Zenvisage~\cite{zenvisagevldb}  
and Qetch~\cite{qetchchi},
two recent 
sketch-based systems for trendline pattern search \agptechreport{(depicted in 	Figure~\ref{fig:baselines})}.
These systems allow users to sketch a pattern on a canvas, 
zoom in and out of the trendline 
to focus on a specific sub-region, 
and apply filtering and smoothing to match trendlines 
at varying granularities. 
While Qetch supports its own custom shape matching algorithm, 
Zenvisage allows users to choose between 
the Euclidean or DTW distance measures 
depending on the task. 
Qetch additionally supports a simple regex (via a \emph{repeat} operator) 
to search for repeated occurrences 
of a sketched pattern. 
We disabled the sketching capability in \ssr 
to isolate the benefits of the novel 
NL and regex query mechanisms 
over sketch. 
\ssrc denotes \ssr with only NL- and regex-based querying mechanisms.
We recruited 24 (14M/10F) participants 
with varying degrees of expertise 
in data analytics via flyers and mass-emails. 
We  employed within-subjects study design between 
\ssrc and each of the baseline tools, 
using two groups of 12 participants each.
Note that by design, each participant
encountered sketch capabilities only once---either in Zenvisage
or Qetch. Participants were free to employ either NL or Regex for \ssrc.
 
 \agptechreport{
\begin{figure}	
	\hspace{-0.2cm}
	\begin{subfigure}{0.61\columnwidth}
		\centerline {
		\hbox{\resizebox{\columnwidth}{!}{\includegraphics{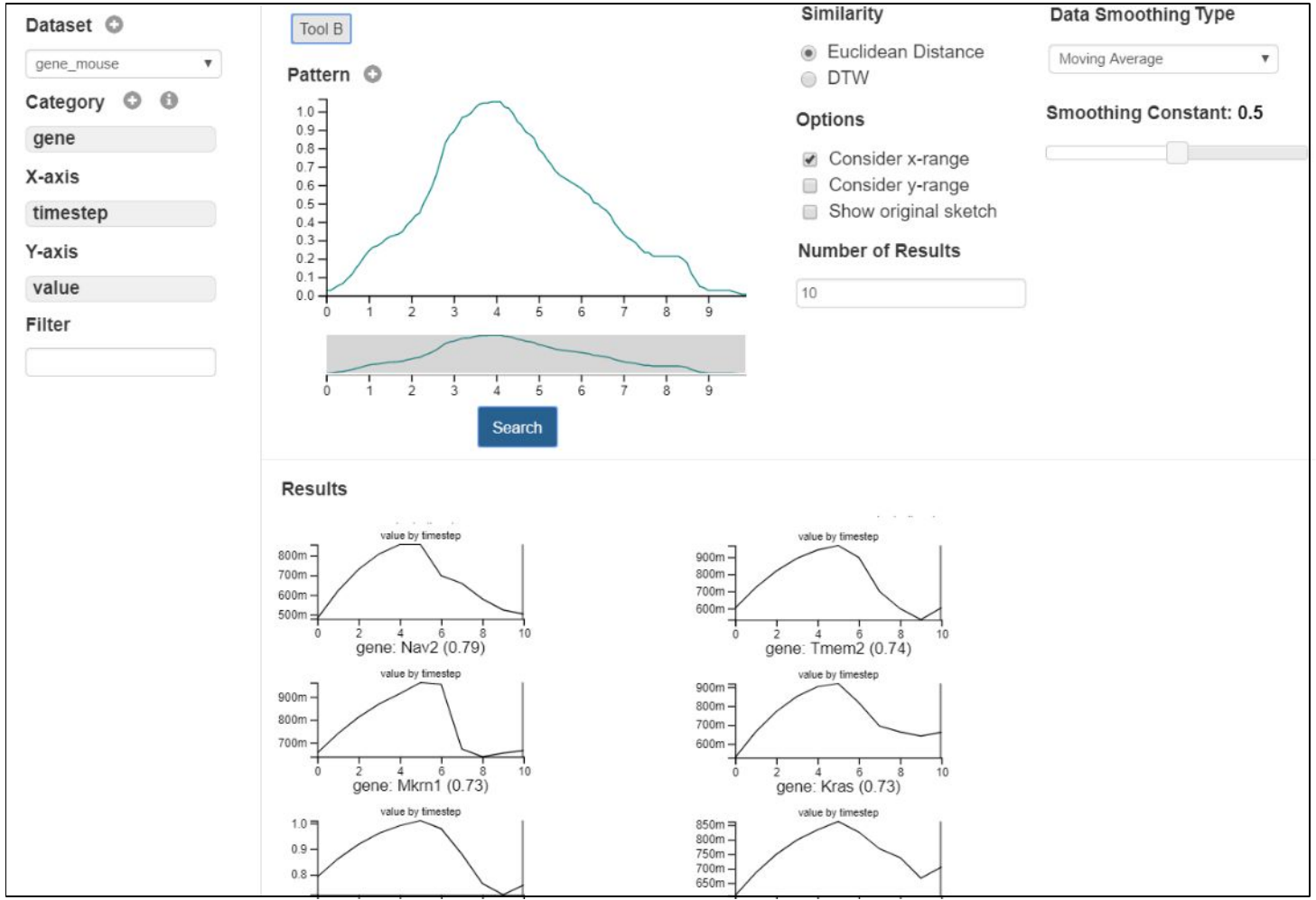}}}}
	\caption{Zenvisage}
	\vspace{-8.5pt}
	\label{fig:baseline_zv}
	\end{subfigure}
	\begin{subfigure}{.39\columnwidth}
		\centerline {
		\hbox{\resizebox{\columnwidth}{!}{\includegraphics{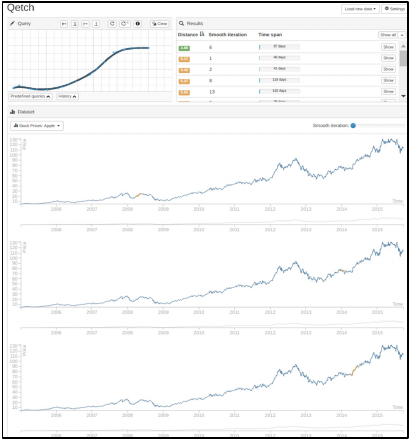}}}}
	\caption{Qetch}
	\vspace{-8.5pt}
	\label{fig:baseline_qetch}
	\end{subfigure}
	\caption{ \smallcaption{Baseline interfaces for user study}}
	\vspace{-5pt}
	\label{fig:baselines}	
\end{figure} 
}



\stitle{Dataset and Tasks.}
Based on the domain case studies from  Section~\ref{sec:intro}, as well as prior work in time series data mining~\cite{ralanamahatana2005mining,ye2009time,fu2011review,olszewski2001generalized,han2011data} and visualization~\cite{2017-regression-by-eye,correll2016semantics,zenvisagevldb,timesearcher,googlecorrelate}, we identified seven categories of pattern matching tasks, as depicted in Table~\ref{tab:study_tasks}. We designed these tasks on two real-world datasets: the Weather and the Dow Jones stock datasets from the UCI repository~\cite{ucirepo} that participants could easily understand and relate with.  Together, the seven tasks spanned both exploratory search as well as targeted pattern-based data exploration, which helped us test the effectiveness of individual interfaces in various settings.


\stitle{Ground Truth.} 
\agppapertext{For each of the tasks, 
three of the authors 
and $20$ Mechanical Turk (mturk) workers,
assigned a score between 
$0$ (worst match) to $5$ (best match) 
to each of the candidate trendlines. 
Each mturk worker was presented with the task description in Table~\ref{tab:study_tasks}, along with a collection of trendlines, each of which they had to rate based on how closely the trendline matched the task description.
We took the average of the ratings 
provided by the three authors and 
mturk workers to be the ground truth. 
We measure the participant's task accuracy 
as the (sum of the ground truth scores of the top K trendlines selected by the participant) $\times$ 100 / (sum of the top K ground-truth scores for the task). K varied between $2$ to $5$ per task.
}

\agptechreport{
For selecting the ground truth, three of the authors independently assigned a score in a range of $5$ (best match) to $0$ (worst match) for each of the trendlines, and filtered out trendlines with average score $<$ 3.0. Next, we leveraged $20$ mturk workers per task to rate each selected trendline in a range of 0-3 (later scaled to $3-5$). 
Each mturk worker was presented with the task description in Table~\ref{tab:study_tasks}, along with a collection of trendlines, each of which they had to rate based on how closely the trendline matched the task description.
 Filtering out noisy trendlines in the first step helped minimize the number of ratings per task, thereby improving the effectiveness of workers.  Finally, we take the average of the scores given by three authors and workers as the ground truth score for a trendline. For a given task, we measure the task-accuracy as (sum of the ground truth scores of the top-K trendlines  selected by the participant) $\times$ 100 / (sum of the top-K ground-truth scores for the task). $K$ varied between $2$ to $5$ per task.
}


\vspace{-5pt}
\subsection{Key Findings}

\begin{table}
	\centering
	\caption{Pattern Matching Tasks} 
	\label{tab:study_tasks}
	\vspace{-12.5pt}
	\resizebox{\columnwidth}{!}{%
		\begin{tabular}{p{2.3cm}|p{8.7cm}}

			{\bf Tasks}  & {\bf Description} \\ \hline
			Exact Trend Match (ET)  &  Find shapes similar to a specific shape, e.g., cities with weather patterns similar to that of NY, stock trends similar to Google's.   \\ \hline
			Sequence Match (SQ)	& Find shapes with similar trend changes over time, e.g., cities with the following temperature trends over time: rise, flat, and fall, stocks with decreasing and then rising trends.  \\ \hline
			Common Trends ( TC) & Summarizing common trends e.g., find cities with typical weather patterns, stock with typical price patterns. \\ \hline
			Sub-pattern Match (SP) & Find frequently occurring sub-pattern, e.g., stocks that depicted a common sub-pattern found in stocks of Google and Microsoft, cities with $2$ peaks in temperature over the year.  \\ \hline
			Width specific Match (WS) & Find shapes occurring over a specific window, e.g., cities with steepest rise or fall in temperature over $3$ months, peaks with a width of $4$ months.  \\ \hline
			Multiple X or Y constraints (MXY) & Find shapes with patterns over multiple disjoint regions of the trendline, e.g., stocks with prices rising in a range of 30 to 60 in march, then falling in the same range over the next month. \\ \hline
		
			Complex Shape Matching (CS) & Find shapes involving trends along specific directions, and occurring over varying duration, e.g., stocks with head and shoulder pattern, cup-shaped patterns, W-shaped patterns. \\ 
		\end{tabular}%
	}
	\vspace{-15pt}
\end{table}

\begin{figure*}
	\vspace{-25.5pt}
	\centering
	\hspace{-0.2cm}
\begin{subfigure}{0.33\textwidth}
	\centerline {
		\hbox{\resizebox{\columnwidth}{!}{\includegraphics{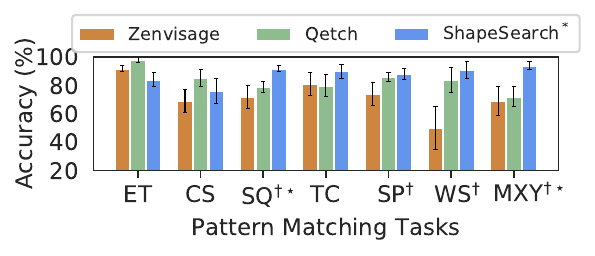}}}}
	\vspace{-5.5pt}
	\caption{\smallcaption{Task Accuracy}}
	\label{fig:study-accuracy}
\end{subfigure}
\hspace{-0.2cm}
\begin{subfigure}{0.33\textwidth}
	\centerline {
		\hbox{\resizebox{\columnwidth}{!}{\includegraphics{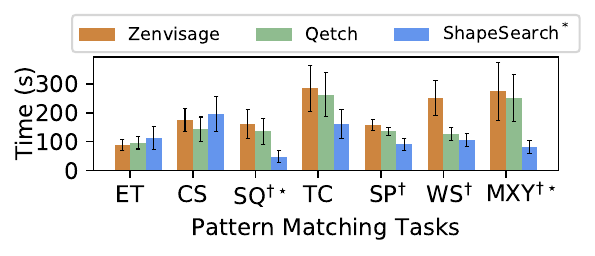}}}}
	\vspace{-5.5pt}
	\caption{\smallcaption{Task Completion Time}}
	\label{fig:study-time}
\end{subfigure}
\hspace{-0.2cm}
\begin{subfigure}{0.33\textwidth}
	\centerline {
		\hbox{\resizebox{\columnwidth}{!}{\includegraphics{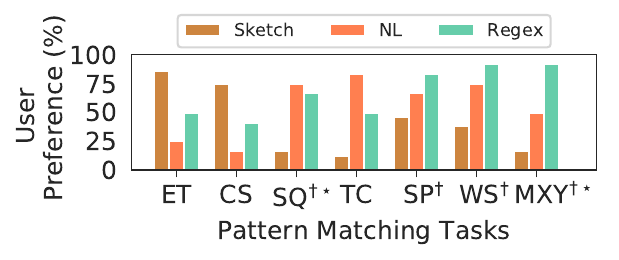}}}}
	\vspace{-5.5pt}
	\caption{\smallcaption{User Preferences}}
	\label{fig:study-preference}
\end{subfigure}
	\vspace{-12.5pt}
\caption{\smallcaption{User study results ($\dagger$  and $\ast$ denote that \ssrc had statistically significant improvements ($\alpha$ = 5\%) relative to Zenvisage and Qetch respectively}}
\label{fig:studyresults}
\vspace{-12.5pt}
\end{figure*}


We describe our key findings below.

\stitle{Overall Task Accuracy and Completion Times.} 
\emph{As depicted in Figure~\ref{fig:study-accuracy} 
and Figure~\ref{fig:study-time}, 
\ssrc helped participants achieve higher accuracy 
and less time overall than Qetch and Zenvisage, 
and in particular for, $5$ out of $7$ tasks; 
however, 
for precise and complex shape matching tasks, \ssrc performed 
worse than baselines due to the lack 
of sketch capabilities.} 
On average across all tasks, \ssrc helped participants 
achieve an accuracy of 87\%---8\% 
more than Qetch and 17\% more than Zenvisage---in 
about 30-40\% less time, 
a significant improvement. 
While Zenvisage and Qetch 
involve less reasoning during query synthesis, 
they often lead to significantly more queries issued 
and manual browsing of trendlines 
for identifying the desired ones. 
\ssrc, on the other hand, 
can accept more fine-grained user queries 
to rank relevant trendlines effectively, 
enabling participants to retrieve more accurate answers with less effort.
In order to better understand the differences between the tools, 
we separately analyze tasks where \ssrc did better 
and worse than the baselines.

\noindent
\textbf{Settings Where \ssrc Wins}. 
Since sketch systems are based on precise matching, 
for sequence and sub-pattern matching tasks (SQ and SP), 
users drew multiple sketches for a given sequence or 
subsequence to find all possible instances. 
\ssrc, however, is effective at automatically 
considering a variety of shapes that satisfy 
the same sequence or subsequence of patterns. 
Similarly, for tasks involving multiple constraints 
along the X and Y axes, or the width of patterns (TC, WS, MXY),
a large majority of the participants 
gave more accurate results in less time with \ssrc. 
\ssrc supports a rich set of primitives for 
users to add multiple constraints to the patterns, 
including searching for patterns over multiple disjoint regions. 
While the users could zoom into a specific region 
of the trendline and sketch their 
desired patterns in the sketch systems, 
these capabilities were not sufficient 
to precisely specify all of the 
constraints at the same time. 
\agptechreport{We believe that supporting visual widgets 
in the baseline tools that internally leverages 
the \ssr primitives could remedy this issue.}


\noindent
\textbf{Settings Where \ssrc Loses}. 
The opposite effect was observed (more time, less accurate with \ssrc)
when finding trendlines exactly similar to a given trendline (ET).
This is understandable given that \ssrc does not possess sketching capabilities, 
which is a perfect fit for this task, and that \ssrc regex 
scoring functions are targeted more towards 
approximate and fuzzy pattern matching.
For complex shapes (CS), 
Qetch performed the best, followed by \ssrc, and then Zenvisage. 
Zenvisage performs the worst because the Euclidean and DTW measures 
used for matching shapes 
are sensitive to distortions in the sketch drawn by users 
for such complex shapes. 
Qetch, on the hand, applies corrections to distortions 
in shapes for better matching. 
For \ssrc the results were mixed. 
We noted that the few participants who over-simplified the shape with fewer patterns (e.g., [\pdown][\pup] for ``cup-shaped'' instead of [\pup][\underline{\pflat}][\pup]) had poorer accuracy  compared to those who used regex appropriately with correct sequence and width constraints.
Overall, we find that complex patterns that involve fuzzy patterns and location constraints are easier to describe using NL and regex than to sketch. In contrast,  complex shapes (e.g., cup shaped) are easier to draw and harder to describe. We believe \ssr with its sketching interface can address the challenges with the latter, and thus support both types of patterns.


\stitle{User Preferences and Limitations.} 
In the end, 
we asked participants to complete a survey 
to gauge their preferences
for the three mechanisms, sketch, NL, and Regex for 
each task. 
(Recall that each participant
encounters a given specification mechanism in only one tool.)
We asked participants to select one or more of the three mechanisms they thought were most suited for each of the tasks they performed. They were allowed to select more than one if they felt multiple mechanisms were helpful.
Figure~\ref{fig:study-preference} depicts the \% of participants who selected the mechanism for each of the tasks.
As depicted in the figure, user 
preferences  
are correlated with their accuracy and completion times: 
most participants preferred the sketch-based interface for precise 
and complex shape-tasks, 
and natural language and regex for other tasks.
When asked about their preferences in general, 
about $62$\% of the participants 
believed that the three interfaces 
integrated together would be most effective, 
$29$\% felt NL and regex together without sketch
would be sufficient for all pattern matching tasks, 
and only $8$\% considered a sketch-based tool as sufficient,
validating our design of a tool that goes beyond sketch capabilities.
\rev{Participant P2 said ``\textit{Almost always, I will go with Tool B} [\ssrc\!\!]. \textit{I know exactly what I am searching [for] and what the tool is going to do, it is much more concise, I feel more confident in expressing my query pattern}". 
About $2/3rd$  of the participants said 
they would opt for regex over natural language or sketch, 
if they had to choose one.  When asked how effective \ssr was in understanding and parsing their natural language queries, the participants gave an average  rating of $3.9$ and when asked how easy it was to learn and apply regular expressions, they gave a rating of $4.4$.
Participant P8 said ``\textit{the concept for visual regex 
by itself is very powerful and could be helpful for most cases in general''}.}

\agptechreport{
\stitle{Other findings.} When asked about the effectiveness of using lines for matching trendlines, the average response was positive with a rating of $4.1$ on a scale of $5$. Participant P4 said ``\textit{Green lines are good, they make me more confident, help me understand trendlines especially [the] noisy ones without me having to spend too much time  parsing signals. I can also see how my [query] pattern was fitted over the trendline ...}''.} 

Finally, participants suggested several improvements to make \ssrc more useful, such as supporting more mathematical patterns; 
automatic regex validation and auto-correction; 
query and trendline recommendations, and 
using different colors for lines that correspond to different patterns (\ssgs) in the \sq.


\vspace{-2pt}
\section{Case Study : Genomics}
\label{sec:casestudy} 

To understand the use of \ssr in a real-world
setting, we 
conducted an 
open-ended evaluation of \ssr 
via a case study with two bioinformatics researchers ($R1$ and $R2$). 
Both researchers are graduate students 
at a major university and perform pattern 
analysis on genomic data on a daily basis 
using a combination of spreadsheets and scripting languages such as R. 
Each session lasted for about $75$ minutes, 
where the researchers explored a popular 
mouse gene dataset~\cite{bult2008mouse} 
that they often analyze as part of their work.

 \subsection{Findings and Takeaways}

\noindent \rom{1}. \emph{ Both participants were able to 
grasp the functionalities of \ssr 
after a 15 minute introduction and demo session without much difficulty}. 
During this session,
the participants appreciated
the ease of pattern search, saying
 ``($R1$) \emph{oh, this feature [searching using combinations of patterns such as up and down] is cool, ... something that we frequently do}'', ``($R2$) \emph{I like that you can change your patterns [queries] that easily, and see the results in no time...}''. Both participants concurred that \ssr could be a valuable tool in a biologist's workflow, and can help perform faster pattern-based data exploration, compared to current R language scripting or spreadsheet approaches. 

\vspace{2pt}
\noindent \rom{2}. \emph{ Using succinct queries, participants could interactively explore a large number of gene groups, depicting a variety of gene expression patterns.}
Both $R1$ and $R2$ were able to query for genes with differential expressions over time.  $R1$ initially issued natural language queries to search for genes that suddenly start expressing themselves at some point, and then gradually stop expressing, i.e., flat, followed by increase, and then gradual decrease, a pattern signifying an effect of external stimulus such as a drug or a treatment. 
\rev{Thereafter, $R1$ was interested in understanding the variations in expression rates, e.g., identifying groups of genes that rise and fall much faster, or where changes are gradual within the same range of values. To search for these patterns, she interactively adjusted the width of patterns, as well as the Y range in her queries via regex. Finally, $R1$ also searched for groups of genes that show similar changes in expression over specific time duration, for finding those that regulated similar cell mechanisms.}

\vspace{2pt}
\noindent 
\rom{3}. \emph{\ssr helped participants validate their hypotheses, and make new discoveries.} 
$R2$ used regex to explore a group of genes that increase with a slope of 45$^\circ$ until a certain point, and then remain high and stable (flat), as well as those with the inverse behavior (ones that start high and then gradually reduce their expression and remain low and flat). Such patterns are typically symbolic of permanent changes (e.g., due to aging) in cell mechanisms, often seen among genes in stem cells. 
 While exploring these patterns, $R2$ discovered two genes,   \textit{gbx2} and \textit{klf5}, in the results panel, that had similar expression patterns within the same range of values, and mentioned that the two genes indeed have similar functionality and are actively being investigated. Next to these two genes, he saw another gene \textit{spry4} with almost similar expression, and hypothesized that the similarity in shape indicates that \textit{spry4} possibly had similar functionalities to \textit{gbx2} and \textit{klf5}, something that is not well-known, and could lead to interesting discoveries if true. \rev{Overall, as can be seen in queries issued by participants, most of the patterns can be expressed using $4$ or fewer number of lines, indicating that it is rare to search for patterns with a large number of \ssgs.}

\vspace{2pt}
\noindent 
\rom{4}. \textit{\ssr helped participants find genes with unexpected or outlier behaviors. During the end of her study, $R1$  mentioned that it is rare to see a gene with two peaks in their expressions within a short window}. However, on searching for this pattern via natural language, she found a gene named ``\emph{pvt1}'' having two peaks within a short time duration of $10$ time points. She found this surprising, and said there could either be some preprocessing error, or some rare activity happening in the cell. She then searched for other unexpected patterns (e.g., three peaks, always increasing).

\agptechreport{
\vspace{2pt}
\noindent
\rom{5}. \textit{Both NL and Regex were equally preferred.}
When asked to compare between NL and regex, $R1$ said she could express most of her queries using natural language, and would use regex only when the pattern is too long, and involves multiple constraints. $R2$, on the other hand, said he would use regex in all scenarios. He believed regex was not significantly difficult to learn, and helped him feel more in control and confident about what he was expressing, and whether the system was correctly inferring and executing his issued queries.

\vspace{2pt}
\noindent
\rom{6}. \emph{Participants faced a few  challenges during exploration.}  They wanted to switch back and forth between queries, so that they do not have to remember and reissue their previous queries.  In addition to better presentation of the  the fitted lines (e.g., coloring), they wanted to understand in more detail  how the scores were computed, and if they could tweak the scoring according to their needs using visual widgets. 
}

\vspace{-5pt}
\section{Related Work}
\label{sec:related}

Our work draws on prior work in visual querying, 
symbolic pattern mining,
as well as natural language interfaces 
for data analytics. 
In Table~\ref{tab:capabilities}, we compare \ssr capabilities 
and expressiveness  
with three representative systems 
from these areas: 
(1)our prior work Zenvisage, 
a general purpose visual querying tool~\cite{zenvisagevldb}, 
(2) Qetch~\cite{qetchchi}, a recent sketch-based system, and 
(3) Shape Definition Language (SDL), 
a symbolic pattern searching language for trendlines. 
At a high-level, \ssr builds on the system
capabilities of visual querying systems as well
as expressiveness of symbolic pattern languages,
while extending both to suit the needs of real
domain users.
Our user study in Section~\ref{sec:userstudy}
compared \ssr with (1) and (2) in terms of usability and effectiveness. 
We summarize key differences with these systems and others below. 
 
Visual querying tools~\cite{wattenberg2001sketching,zenvisagevldb,googlecorrelate,ryall2005querylines, muthumanickam2016shape} 
help search for visualizations with a desired shape 
by taking as input a sketch of that shape. 
Most of these tools perform 
precise point-wise matching using measures 
such as Euclidean distance or DTW.  
A few tools such as TimeSearcher~\cite{timesearcher}  
let users apply soft or hard constraints 
on the $x$ and $y$ range values via boxes or query envelopes, 
but do not support mechanisms for specifying shape 
primitives beyond location constraints.
Qetch improves upon these systems 
by supporting a custom similarity metric 
that is robust to distortions in the user sketch, 
in addition to supporting a ``repeat'' operator for 
finding recurring patterns. 
However, as depicted in Table~\ref{tab:capabilities},
and discussed in Section~\ref{sec:userstudy}, 
Qetch and other visual querying tools 
have limited expressiveness 
when it comes to fuzzy pattern match
needs. 
Furthermore, \ssr introduces
a novel algebra that improves
extensibility by acting as a common
``substrate'' for various input mechanisms, 
along with an optimization engine that efficiently matches patterns
against a large collection of 
trendlines.

Symbolic sequence matching papers approach
the problem of pattern matching by employing
offline computation to chunk trendlines
into fixed length blocks, encoding each block
with a symbol that describes the pattern in that block~\cite{psaila1995querying, faloutsos1994fast,lin1995fast,shatkay1996approximate,garofalakis1999spirit}. 
The most relevant one of these papers is on the 
Shape Definition Language (SDL)~\cite{psaila1995querying}, which
encodes each block using ``up'', ``down'', and ``flat'' patterns,
much like \ssr, and supports a language
for searching for patterns based on their sequence 
or the number of occurrences. 
Since SDL operates on pre-chunked-and-labeled
trendlines, the problem is one of
matching regular expressions
against string sequences (one per pre-labeled
trendline).
Therefore, SDL cannot rank these trendlines,
instead only returning 
a boolean score for whether the pattern matches
the string sequence.
This limits the expressiveness of SDL (Table~\ref{tab:capabilities}),
especially when the patterns are more complex,
as well as when they don't align perfectly well
with the boundaries of the blocks used for chunking.
Moreover, since the trendlines are pre-labeled and indexed,
SDL does not support on-the-fly pattern matching 
where the same trendline can change shapes 
based on filters or aggregation constraints. 
\ssr, on the other hand, adopts a more online query-aware 
ranking of trendlines 
without requiring precomputation, 
and is thus more suited for ad-hoc data exploration scenarios.


\rev{There are a few visual time series exploration tools such as Metro-Viz~\cite{eichmann2019visual} and ONEX~\cite{neamtu2017interactive} that support other analytics tasks such as anomaly detection and clustering.}
There is also a large body of work on keyword- and natural language-based interfaces for querying databases~\cite{li2014constructing} and generating visualizations~\cite{gao2015datatone,eviza}. However, since the underlying shape query algebra in \ssr is different from SQL, parsing and translation strategies from existing work cannot be easily adapted.

\begin{table}
	\centering
		\vspace{-12.5pt}
	\caption{\small \ssr vs.~related systems capabilities} 
	\label{tab:capabilities}
	\vspace{-12.5pt}
	\resizebox{0.9\columnwidth}{!}{%
		\begin{tabular}{p{3.35cm} c c c p{2.5cm}}
			Aspect  & Zenvisage & Qetch & SDL & \small{\ssr}  \\ \hline \hline
			\textbf{System Capabilities} & &  &  &  \\ 
			Precise Pattern & \cmark \cmark  & \cmark \cmark &   \xmark &  \cmark \cmark  \\ 
			Fuzzy Pattern &  \xmark & \cmark & \cmark \cmark  &  \cmark \cmark  \\ 
			\pbox{2cm}{Specification} & sketch &  sketch &  regex & sketch, NL, Regex\\ 
			Auto Smoothing	& \xmark & \cmark \cmark & \cmark &  \cmark \\ 
			Algorithm &  ED, DTW & Custom & Custom & Custom \\ 
		    Ad hoc Patterns & \cmark &  \cmark &  \xmark &  \cmark \cmark \\ 
			Normalization  &  \cmark &  \cmark  \cmark &  \xmark &  \cmark (z-score) \\ 
			Indexing Needed & \cmark &  \cmark & \xmark  & \cmark\\ 
			Scalability  & \cmark \cmark &  \cmark &  \cmark  &  \cmark \cmark \\ 
			Extensibility & \xmark & \xmark &  \xmark &  \cmark \cmark \\  \hline
			\textbf{Query Expressivity} & &  &  & \\  
			\pbox{3.5cm}{Range Constraints } & \cmark & \cmark & \xmark & \cmark \cmark  \\ 
			\pbox{3.5cm}{Sub-Pattern  Matching}  & \cmark & \cmark & \cmark & \cmark \cmark  \\ 
			\pbox{3.5cm}{Sequence Matching}  & \xmark & \xmark & \cmark \cmark & \cmark \cmark   \\ 
			\pbox{3.5cm}{Width Selection}  & \xmark & \xmark & \xmark & \cmark \cmark  \\ 
			\pbox{3.5cm}{Multi- X or Y Constraints}   & \xmark & \xmark & \xmark & \cmark  \\ 
			\pbox{3.5cm}{Quantifiers}  & \xmark & \cmark (repeat) & \cmark \cmark & \cmark \cmark  \\ 
			\pbox{3.5cm}{Iteration}  & \xmark & \xmark & \xmark & \cmark  \\ 
			\pbox{3.5cm}{Nesting}  & \xmark & \xmark & \xmark & \cmark  \\ 
			\pbox{3.5cm}{Back/Forward Reference} & \xmark& \xmark & \xmark & \cmark  \\
		\end{tabular}%
	}
	\vspace{-13pt}
\end{table}

\vspace{-7pt}

\section{Conclusion}
\label{sec:conclusion}
We presented \ssr, an end-to-end pattern search system, 
providing flexible mechanisms 
for domain experts to effortlessly 
and efficiently search for trendlines  
with desired shapes. 
We introduced \sq, which forms the core of \ssr, 
and helps express a large variety of 
patterns with a minimal set 
of primitives and operators, as well as 
an execution engine that enables interactive
pattern matching on a large collection of visualizations.
Our user study, case study with genomics researchers,
along with performance experiments
demonstrate the efficiency, effectiveness, 
and usability of \ssr.
\ssr is a promising step
towards accelerating the search for insights in data,
while catering to the needs of expert and novice
programmers alike.


\noindent {\bf Acknowledgments.} We thank the anonymous \agptechreport{SIGMOD 2020} reviewers for their valuable feedback. We acknowledge support from grants IIS-1652750 and IIS-1733878 awarded by the National Science Foundation, grant W911NF-18-1-0335 awarded by the Army, and funds from Facebook, Adobe, Toyota Research Institute, Google, and the Siebel Energy Institute. The content is solely the responsibility of the authors and does not necessarily represent the official views of the funding agencies and organizations.

\agppapertext{\vspace{-8pt}}
{\scriptsize
	\bibliographystyle{abbrv}
	\bibliography{main}
}

\agptechreport{
\appendix 
\section{Appendix}
Here, we explain our methodology for crowdsourcing natural language-based pattern queries in trendlines. We use crowd-sourced queries for two purposes. First, we analyze them for 
characterizing trendline patterns, that we explain in Section~\ref{sec:intro}. Second, we use it for training a parser for automatically translating natural language queries to \sqs. We explain the features and translation steps in Section~\ref{sec:translation}.

\subsection{Crowd Study Methodology}
\label{sec:mturk}
We conducted the crowd study using Amazon Mechanical Turk, where we asked workers to describe patterns in trendlines using English language sentences. We describe the steps below.

We first manually collected a total of 50 trendlines (called anchor trendlines) with varying patterns from $4$ datasets: Worms, 50 words, Haptics, Weather from the UCI Machine repository~\cite{ucirepo}.  We mixed each of the anchor trendlines with $19$ other trendlines from the same dataset, to create 50 collections of $20$ trendlines each.

Using these collections, we conducted a Mechanical Turk study with a total of about $265$ workers. In order to ensure good quality, each worker was selected through a pre-screening HIT that tested  their basic English language fluency and the ability to write reasonably meaningful English sentences.

Each worker was presented with an interface depicting $20$ trendlines corresponding to one of the $50$ collections. We highlighted the anchor trendline by bordering it with a green-colored box. Moreover, all trendlines had  X and Y axis values labeled. We asked workers to describe the pattern in the anchor trendline using an English sentence. In addition, we suggested that their description should be helpful in locating the anchor trendline if it was not highlighted. Workers had to write the English description in a textbox at the top of the interface.

After filtering out responses that did not address the task, there were a total of $250$ English sentences, with about $5$ English sentences on average for each of the anchor trendlines.

\stitle{Analysis.} In addition to manual inspection, we performed text-analysis on collected sentences to understand some of the frequent words as well as sentence structure used by workers for describing patterns. We noticed that a large majority ($>$ $80$\%) of the sentences included either ``increasing",`` decreasing", ``flat" or their synonyms. Moreover, whenever there were multiple occurrences of these words in the same sentence, they were frequently separated by ``and'', ``and then'', ``next'', ``,''.  Many of the sentences also included ordering words such as first, second, or third. 
While a large majority of the workers did not provide details on the X and Y range values of individual patterns in the query, those who did mostly mentioned the start and end locations of the individual patterns. Overall, more than $98$\% of the sentences included less than $20$ words, and $< 5$ patterns per sentence. We summarize the key characteristics of collected queries in Section~\ref{sec:intro}.

\stitle{Labeling for NL to \sq translation .}  We also used the collected sentences for training a conditional random field (CRF) model for translating natural language queries to \sqs. In order to do so, we manually annotated the words in the collected queries with primitives and operators supported in the \sq algebra. We used the annotated queries for training a conditional random field (CRF) model for translating natural language queries to \sqs. We explained the features and translation steps in Section~\ref{sec:translation}. On $5$-fold cross-validation over these queries, the model had an $F1$ score of $81\%$ ($precis on=73\%$, $recall=90\%$), showing that the structure and key constructs (e.g., primitives and operators in \sq algeabra) in natural language-based pattern queries have high degree of predictability. 

\rev{
\subsection{Approximate Matching using Sketch}
\label{sec:sketchapprox}
}

\rev{
In this section, we provide more details on how a sketch is translated to regex for approximate matching. This process consists of two steps:
1) converting a sketch to a sequence of minimal number of line segments, and 2) constructing a regex query using the slopes of the line segments. We describe each of these steps below.
}

\rev{
	
\stitle{1. Converting sketch to a sequence of lines.} Given an user-drawn sketch, \ssr approximates it using as fewer number of lines as possible. However, too few lines can often lead to a  poor approximation of the sketch, e.g., approximating a bell-shaped sketch with a single line segment. In order to avoid this, we minimize the number of lines with a constraint that the approximation error is within a specific threshold $e$. For doing so, we take as input a smoothing granularity, $s$, between $0$ and $1$, that users can vary via a slider  (Figure~\ref{fig:interface}-2a). Higher the smoothing granularity, the fewer the number of lines needed to approximate the sketch, and vice-versa. Internally, smoothing is translated to a $R^2$ error~\cite{r2wiki} threshold, $e$, as $e = 1 - s.$ We note that the problem of finding the minimum number of lines within an error threshold $e$ is a well-studied problem in time series.
}

\rev{
\begin{problem}
Given a time series T, find the minimal line approximation of T such that the combined $R^2$ error for all lines does not exceed $e$. 
\end{problem}
}

\rev{
If we knew the minimal number of lines in advance,  the problem can be optimally solved using a dynamic programming algorithm~\cite{terzi2006efficient}.
However, it is difficult to know the minimum number of lines in advance, and thus we use another top-down segmentation  algorithm~\cite{keogh2004segmenting} that has been well-studied in time series.
The algorithm starts with a single line approximation, and  recursively segments lines into more lines until the $R^2$ error is below $e$. For choosing the point for segmentation, e.g., from a single line to two lines, the top-down algorithm considers every point for segmentation and chooses the one that leads to the maximum reduction in $R^2$ error after segmentation. The time complexity of the algorithm for a trendline with $n$ points and $K$ number of lines is  $O(n^2 \times K)$. 
}

\rev{
\stitle{2. Constructing a  regex query.} After approximating the sketch with lines, \ssr constructs a regex query using the slopes of the lines. Formally, given $K$ lines with slopes $\theta_1, \theta_3, \theta_3,..., \theta_k$, \ssr constructs the following regex:
[{\color{olive} p= $\theta_1$}]\colorconcat[{\color{olive} p= $\theta_2$}] ... \colorconcat[{\color{olive} p= $\theta_k$}]. 
After translation, the regex representation of the sketch is shown to the user 
for validation in the correction panel (Figure~\ref{fig:interface} Box 3) 
The validated query is finally optimized and executed,  and the top visualizations that best match the \sq are presented in the results panel (Figure~\ref{fig:interface} Box 4).
}

}

\end{document}